\documentclass[preprint,aps,pra,showpacs,floatfix,superscriptaddress]{revtex4-1}
\usepackage{graphicx}
\usepackage{feynmp}
\usepackage{amsmath,amsfonts,amssymb,verbatim,float}
\usepackage{epsf}
\usepackage{bm}
\usepackage{bbm}
\usepackage[usenames]{color}	
\usepackage{colortbl}
\usepackage{hyperref} 
\usepackage{indentfirst}
\usepackage{siunitx} 
\usepackage{calrsfs}
\usepackage{calrsfs}
\usepackage{ulem} 
\usepackage{subfigure}
\newcommand{\balpha}{\boldsymbol{\alpha}}

\graphicspath{{graph/}}
\newcommand{\be}{\begin{eqnarray}}
\newcommand{\ee}{\end{eqnarray}}

\newcommand{\veps}{\varepsilon}

\newcommand{\bnabla}{\bm{\nabla}}

\newcommand{\bfp}{{\bf p}}

\newcommand{\bfr}{{\bf r}}

\newcommand{\aZ}{\alpha Z}
\newcommand{\mub}{\mu_0}

\newcommand{\ket}[1]{|#1\rangle}

\newcommand{\matrixel}[3]{\langle #1 | #2 | #3 \rangle}

\newcommand{\gdirac}{g_{\rm D}}
\newcommand{\dgint}{\Delta g_{\rm int}}
\newcommand{\dgqed}{\Delta g_{\rm QED}}

\newcommand{\dgnuc}{\Delta g_{\rm nuc}}

\newcommand{\dgvp}{\Delta g_{\rm VP}}
\newcommand{\dgvpel}{\Delta g_{\rm VP}^{\mathrm{el}}}
\newcommand{\dgvpm}{\Delta g_{\rm VP}^{\mathrm{m}}}
\newcommand{\dgirr}{\Delta g_{\rm irr}}
\newcommand{\dgirrlo}{\dgirr^{(0)}}
\newcommand{\dgirrfo}{\dgirr^{(1)}}
\newcommand{\dgirrm}{\dgirr^{(m)}}

\newcommand{\dgvr}{\Delta g_{\rm vr}}
\newcommand{\dgvrlo}{\dgvr^{(0)}}
\newcommand{\dgvrfo}{\dgvr^{(1)}}
\newcommand{\dgvrm}{\dgvr^{(m)}}

\newcommand{\hmm}{{H}_M^{\textrm{magn}}}
\newcommand{\hml}{{H}_M}
\newcommand{\dgrecfull}{\Delta g_{\rm rec}}

\newcommand{\dgrecmgn}{\Delta g_{\rm rec}^{\rm magn}}
\newcommand{\dgrecnonmgn}{\Delta g_{\rm rec}^{\rm non-magn}}
\usepackage{color}
\definecolor{BLUE}{rgb}{0.0,0.0,1.0}

\usepackage{soul}
\soulregister\cite7
\soulregister\ref7

\begin{document}
\thispagestyle{empty}
\title{
$g$ factor of the $2p_j$ excited states in lithium-like ions
}
\author{D. V. Zinenko}
\affiliation{Department of Physics, St. Petersburg State University, Universitetskaya nab. 7/9, 199034 St. Petersburg, Russia}
\author{V. A. Agababaev}
\affiliation{School of Physics and Engineering, ITMO University, Kronverkskiy pr. 49, 197101 St. Petersburg, Russia}
\author{D. A. Glazov}
\affiliation{School of Physics and Engineering, ITMO University, Kronverkskiy pr. 49, 197101 St. Petersburg, Russia}
\affiliation{Petersburg Nuclear Physics Institute named by B. P. Konstantinov of National Research Centre “Kurchatov Institute”, Gatchina, 188300 Leningrad District, Russia}
\author{A. V. Malyshev}
\affiliation{Department of Physics, St. Petersburg State University, Universitetskaya nab. 7/9, 199034 St. Petersburg, Russia}
\affiliation{Petersburg Nuclear Physics Institute named by B. P. Konstantinov of National Research Centre “Kurchatov Institute”, Gatchina, 188300 Leningrad District, Russia}
\author{A.~D.~Moshkin}
\affiliation{Department of Physics, St. Petersburg State University, Universitetskaya nab. 7/9, 199034 St. Petersburg, Russia}
\affiliation{School of Physics and Engineering, ITMO University, Kronverkskiy pr. 49, 197101 St. Petersburg, Russia}
\author{E. V. Tryapitsyna}
\affiliation{School of Physics and Engineering, ITMO University, Kronverkskiy pr. 49, 197101 St. Petersburg, Russia}
\affiliation{Petersburg Nuclear Physics Institute named by B. P. Konstantinov of National Research Centre “Kurchatov Institute”, Gatchina, 188300 Leningrad District, Russia}
\author{A. V. Volotka}
\affiliation{School of Physics and Engineering, ITMO University, Kronverkskiy pr. 49, 197101 St. Petersburg, Russia}
%
\begin{abstract}
%
Relativistic calculations for the $g$ factor of the lowest excited states $2p_{1/2}$ and $2p_{3/2}$ of lithium-like ions over a wide range of the nuclear charge numbers $Z=10-92$ are presented. Interelectronic interaction is considered within the perturbation theory up to the second order. One-loop QED contributions are calculated to all orders in $\alpha Z$. Leading contributions of the nuclear recoil effects are taken into account. The quadratic and cubic terms in magnetic field are considered as well. A set of screening potentials is used in all calculations to estimate the unknown higher-order correlation effects.
\end{abstract}
%

\maketitle
%
%
\section{Introduction}
\label{sec:intro}
The study of the $g$ factor of highly charged ions has made considerable progress in recent years \cite{Shabaev:2015:031205, Glazov:2023:A}. Joint theoretical and experimental results for hydrogen-like ions have led to the most precise measurement of the electron mass to date \cite{sturm:2014:467,CODATA:2025:025002}. It is reasonable to expect that further improvements in both theory and experiment will allow an independent determination of the fine structure constant $\alpha$ \cite{Shabaev:2006:253002, Yerokhin:2016:100801}, facilitate tests of QED beyond the Furry picture in the strong coupling regime \cite{Shabaev:2017:263001, Malyshev:2017:765, Shabaev:2018:032512}, as well as measurement of nuclear magnetic moments and nuclear radii \cite{Quint:2008:032517, Yerokhin:2011:043004, Sailer:2022:479}. Recently, it has been shown that the ground-state $g$-factor measurement in hydrogen- and lithium-like thorium-229 ions can provide determination of the lifetime of the isomeric state \cite{shabaev:2022:043001}, which is crucial for the development of nuclear clock \cite{Peik:2003:181, Campbell:2012:120802} and the nuclear transition laser \cite{Tkalya:2011:162501}. Accurate $g$-factor values are also in demand in astrophysics. A technique based on magnetic dipole (M1) transitions in highly charged ions has been proposed for measuring magnetic field components in the solar corona \cite{Schiffmann:2021:186}.

To date, a number of high-precision experiments have been conducted to determine the bound-electron $g$ factor in highly charged ions. $g$-factor measurements in light hydrogen-like ions have achieved an accuracy of $10^{-10}$--$10^{-11}$ \cite{Haffner:2000:5308, Verdu:2004:093002, Sturm:2011:023002, Sturm:2013:030501, Heisse:2023:PRL, Morgner:2023:53}. Further investigations of lithium- and boron-like ions were additionally stimulated by considering the so-called specific differences, which circumvent the problem of finite nuclear size and significantly improve the accuracy of the comparison between theory and experiment \cite{Shabaev:2002:062104, Volotka:2014:023002}. For lithium-like silicon \cite{Wagner:2013:033003, Glazov:2019:173001} and calcium \cite{Kohler:2016:10246} the accuracy achieved the level of $10^{-9}$--$10^{-10}$. Within the ALPHATRAP project, a high-precision measurement was carried out for the ground state of boron-like argon ${}^{40}\mathrm{Ar}^{13+}$ \cite{Arapoglou:2019:253001}. 

The technique of all the experiments mentioned above is suitable only for the ground state of highly charged ions. Measurements of Zeeman splitting in excited states have thus far been limited to light systems such as helium and lithium. However, a precise measurement of the fine structure interval in boron-like argon \cite{SoriaOrts:2007:052501} facilitated the determination of the $g$ factor of the lowest excited $2p_{3/2}$ state with an accuracy of $10^{-3}$. This is not yet sufficient to test for non-trivial contributions. However, recent experimental developments allow for higher precision for $g$ factors of excited states. Improved fine-structure laser spectroscopy within the ALPHATRAP project yielded an order-of-magnitude more precise $g$ factor of the $2p_{3/2}$ state in boron-like argon~\cite{Egl:2019:123001}. An improvement further to the level of $10^{-6}$ was achieved using quantum logic spectroscopy~\cite{Micke:2020:60}. This method enables future measurements of the Zeeman splitting of different states of highly charged ions, including excited states, without being restricted to boron-like ions. The first results for carbon-like calcium $\mathrm{Ca}^{14+}$ have been published recently \cite{spiess:25}.

Theoretical calculations of the Zeeman effect for excited states have so far been carried out mainly only in the lowest relativistic approximation \cite{Guan:1998:120, Yan:2001:5683, Yan:2002:1885, Yan:2002:022502}. Generally, the accuracy of these calculations does not exceed the level of $10^{-4}$, with the only exception being the $2p_{3/2}$ state in boron-like ions. Relativistic calculations of the correlation and QED effects have been carried out, yielding results with an uncertainty of $10^{-6}$ for the range of $Z = 10 - 20$ \cite{Glazov:2013:014014, Maison:2019:042506, Agababaev:2020:143}. $g$ factor of boron-like tin has been calculated recently in view of the corresponding measurement~\cite{Morgner:2025:123201}. Note that the calculations performed by the MCDF method \cite{Verdebout:2014:1111, Marques:2016:042504} were inaccurate due to incomplete consideration of the contribution of negative energy states.

In this work, we present the first relativistic calculations of the $g$ factor of excited states in lithium-like ions. The interelectronic interaction correction is considered in the framework of perturbation theory. The first-order $1/Z$ term is obtained within a rigorous QED approach, i.e., to all orders in $\alpha Z$. The second-order contribution is calculated in the Breit approximation, taking into account excitations to the negative spectrum. The one-loop QED contributions, self-energy and vacuum polarization, are calculated to all orders in $\alpha Z$. The currently known terms of the $\alpha Z$-expansion are used to calculate the two-loop corrections. Nuclear recoil effects have also been taken into account using relativistic effective operators. All contributions are calculated with the screening potentials included in the Dirac equation (so-called extended Furry picture) to allow for partial consideration of higher-order contributions. The spread of total values obtained with different potentials serves as an estimate of the uncalculated contributions, as the exact result should not depend on the initial approximation. In this paper, we employ the core-Hartree potential and the family of $x_{\alpha}$-potentials (Kohn-Sham, Dirac-Hartree, and Dirac-Slater potentials, see, e.g., \cite{Sapirstein:2002:042501}). As a result, we have obtained theoretical values of the $g$ factor of the excited states $(1s)^2 2p_{1/2}$ and $(1s)^2 2p_{3/2}$ in lithium-like ions for $Z = 10$--$92$, with an accuracy of $10^{-6}$.

For the states under consideration, it is also important to take into account the contributions of nonlinear Zeeman effects, which are enhanced for closely spaced levels admixed by an external magnetic field. Previous calculations of these effects for boron-like ions \cite{vonLindenfels:2013:023412, Glazov:2013:014014, Agababaev:17:NIMB, Varentsova:2017:80, Varentsova:2018:043402} included correlation effects. Recently, QED correction has been evaluated for $1s$, $2s$, and $2p_{1/2}$ states~\cite{Agababaev:2025}. For the present theoretical accuracy of the Zeeman splitting, the leading-order calculation is sufficient. Note, that these effects are important for interpretation of high-precision experimental data~\cite{Arapoglou:2019:253001, Egl:2019:123001}.

The paper is structured as follows. Section \ref{sec:basic} presents the fundamental expressions for the bound-electron $g$ factor in few-electron ions. Each correction to the $g$ factor is then discussed individually: interelectronic interaction in Subsection \ref{sec:me}, QED corrections in Subsection \ref{sec:qed} and nuclear recoil effects in Subsection \ref{sec:rec}. Section \ref{sec:results} presents the numerical results.


Relativistic units ($\hbar = 1$, $c = 1$, $m_e = 1$) and the Heaviside charge unit [$\alpha = e^2/(4\pi)$, $e<0$] are used throughout the paper.
%
%
\section{Theory}
\label{sec:basic}
We consider the Zeeman effect, i.e., the splitting of atomic energy levels due to a weak, homogeneous external magnetic field applied to an atom with a spinless nucleus. 
The energy of the state $|a\rangle$ can be expanded in a power series of the magnetic field strength 
\begin{equation}
\label{eq:pt1}
 E = E^{(0)} + E^{(1)} + E^{(2)} +  E^{(3)} + \dots,
\end{equation}
where $E^{(0)}$ is the unperturbed energy of the state $|a\rangle$.
The first order energy shift is expressed in terms of the electronic $g$ factor:
\be
    E^{(1)}  = -\dfrac{e}{2} g m_j B \,,
  \label{eq:e_shift} 
\ee 
where $m_j$ is the projection of the total angular momentum $j$ along the magnetic field direction.
%
In the case of one electron over the closed shells and a spinless nucleus, $m_j$ is determined by the valence electron state $\ket{a} = \ket{j_a m_a}$, with the angular momentum $j_a$ and its projection $m_a$. In the $(1s)^2\, 2p_{1/2}$ and $(1s)^2\, 2p_{3/2}$ states of a lithium-like ion, these correspond to the $2p_{1/2}$ and $2p_{3/2}$ states, respectively.

The quadratic and the cubic energy shifts can be factorized using dimensionless quantities $g^{(2)}$ and $g^{(3)}$, respectively:
\begin{equation}
    E^{(2)} =  (-\dfrac{e}{2} B)^2 g^{(2)}(m_j) \, ,
    \label{eq:E2} 
\end{equation}

\begin{equation}
    E^{(3)} =  (-\dfrac{e}{2} B)^3 g^{(3)}(m_j) \, ,
    \label{eq:E3}
\end{equation}
The dependence of the quadratic Zeeman effect on $m_j$ is more complex and cannot generally be factorized.


The interaction operator of the bound electron with the external magnetic field $\boldsymbol{B}$ can be written in the following form
\be
V_{\rm m} = -e\bm{\alpha}\cdot \bm{A(r)}=-\frac{e}{2}BU\,,
\label{eq:V_magn} 
\ee
where, without loss of generality, $B$ is assumed to be oriented along the $z$-direction, and $U = [\bfr\times \balpha]_z$, with $\balpha$ denoting the vector of Dirac matrices.
The one-electron wave function obeys the Dirac equation,
\be
\label{eq:Dirac}
  h^{\rm D} \ket{a} = \varepsilon_a \ket{a}
\,,\qquad
  h^{\rm D} = -i \balpha \cdot \bnabla + \beta + V(r)
\,,
\ee
where the binding potential $V(r)$ includes the nuclear potential and optionally some effective screening potential. 
Within the independent-electron approximation, the first-order energy shift $E^{(1)}$ is found as the expectation value of $V_{\rm m}$ with $\ket{a}$, which yields the following expression for the $g$ factor,
\be
\label{eq:gbreit}
 g^{(0)} = \dfrac{1}{m_a}\langle a \vert U \vert a \rangle 
\,.
\ee
For the pure Coulomb nuclear potential, the $g$ factor is denoted as $g^{(0)}_\mathrm{C}$. In the case of a point nucleus, $g^{(0)}_\mathrm{C}$ is known analytically (we denote it as $\gdirac$) and, for the states $2p_{1/2}$ and $2p_{3/2}$ of interest to us, is given by the following formulae \cite{Breit:1928:649}:
\be
\gdirac (2p_{1/2}) = \dfrac{2}{3}\bigg[\sqrt{2\big(1+\sqrt{1 - (\aZ)^2\big)}} - 1 \bigg] = \frac{2}{3} - \frac{1}{6}(\aZ)^2 - \dots
\,,
\ee

\be
\gdirac (2p_{3/2}) = \dfrac{4}{15}\bigg[2\sqrt{4 - (\aZ)^2} + 1 \bigg]  = \frac{4}{3} - \frac{2}{15}(\aZ)^2 - \dots
\ee

The total $g$-factor value comprises $\gdirac$ and various corrections,
\be
\label{eq:gtotal}
  g = \gdirac + \dgint + \dgqed + \dgnuc
\,.
\ee
Here, $\dgint$ is the interelectronic-interaction correction, $\dgqed$ is the QED correction, $\dgnuc$ stands for the nuclear recoil effect. As in our previous work \cite{Zinenko:2023:032815}, all calculations are performed within the extended Furry picture, i.e., with inclusion of the screening potential in the Dirac equation.

We also want to take into account the corrections related to the quadratic \eqref{eq:E2} and cubic \eqref{eq:E3} energy shifts. Due to the symmetry regarding the sign of the projection of magnetic quantum number $m_j$, the quadratic effect has no impact on the Zeeman splitting in the ground states of highly charged ions. However, for the excited states $2p_{1/2}$ and $2p_{3/2}$ in lithium-like ions, this effect becomes observable \cite{vonLindenfels:2013:023412}. For example, the quadratic Zeeman effect provides a small shift of the fine structure levels, which was taken into account to obtain the most accurate current experimental value of the fine structure transition energy in boron-like argon \cite{Egl:2019:123001}. Additionally, the theoretical results of previous studies on the cubic Zeeman effect \cite{Varentsova:2018:043402} were used to obtain the experimental value of the $g$ factor of boron-like argon in the ALPHATRAP experiment. The corresponding contribution to the $g$ factor was obtained with an accuracy of $10^{-9}$ \cite{Arapoglou:2019:253001}. 

The Zeeman quadratic effect $g^{(2)}$ is calculated as the sum over the spectrum

\be
  g^{(2)} = {\sum_{n}}'
    \frac{\matrixel{a}{U}{n}
          \matrixel{n}{U}{a}}
         {\veps_a-\veps_{n}}.
\ee

The expression for the third-order contribution $g^{(3)}$ is written in the following form

\be
  g^{(3)} = \sideset{}{'}\sum_{n_1, n_2}
    \frac{\matrixel{a}{U}{n_1}
          \matrixel{n_1}{U}{n_2} \matrixel{n_2}{U}{a}}
         {(\veps_a-\veps_{n_1})(\veps_a-\veps_{n_2})} - {\sum_{n}}'
    \frac{\matrixel{a}{U}{n}
          \matrixel{n}{U}{a}}
         {(\veps_a-\veps_{n})^2} \matrixel{a}{U}{a}.
\ee

This effect was previously calculated for the ground state of hydrogen-like, lithium-like, and boron-like ions \cite{Varentsova:2017:80}.


%
%
\subsection{Interelectronic-interaction correction} 
\label{sec:me}
In the framework of bound-state QED perturbation theory, the interelectronic-interaction contribution $\Delta g_{\rm int}$ can be expressed as
\be
\label{eq:g_int}
  \Delta g_{\rm int} &=& \Delta g^{(0)}_{\rm int} + \Delta g^{(1)}_{\rm int} + \Delta g^{(2)}_{\rm int}
\,,
\ee
where $\Delta g^{(i)}_{\rm int}$ denotes the $i$th-order correction in $\alpha$. Specifically, $\Delta g^{(1)}_{\rm int}$ and $\Delta g^{(2)}_{\rm int}$ correspond to the one- and two-photon exchange contributions to the bound-electron $g$ factor, respectively. The zeroth-order term $\Delta g^{(0)}_{\rm int}$ is just the difference between the one-electron values in the extended and original Furry pictures,
\be
\label{eq:g_scr}
  \Delta g^{(0)}_{\rm int} &=& g^{(0)} - g^{(0)}_\mathrm{C}
\,.
\ee
Currently, there are several methods to derive formal expressions from the first principles of QED: the two-time Green's function method \cite{shabaev:2002:119}, the covariant-evolution-operator method \cite{lindgren:2004:161}, and the line profile approach \cite{andreev:2008:135}. 
The first-order term $\Delta g^{(1)}_{\rm int}$ is calculated strictly within the framework of the QED approach, i.e., to all orders in $\alpha Z$.
The corresponding expression for the $\Delta g^{(1)}_{\rm int}$ can be found in \cite{Zinenko:2023:032815}.
%
%
The QED equations for the two-photon exchange correction $\Delta g^{(2)}_{\rm int}$ are also presented in \cite{Zinenko:2023:032815}. At present, it is rather difficult to perform second-order calculations strictly within the framework of the QED theory, so approximate methods are used. In this work, the calculations are carried out in the Breit approximation. The corresponding expressions for $\Delta g^{(2)}_{\rm int}$ in this approximation can be obtained from the QED equations by replacing the interelectronic interaction operator $I(\omega)$ with $I_B$, defined as
\begin{equation}
  I_{\rm B}(r_{12}) = \alpha \left( \dfrac{1}{r_{12}} 
    - \frac{\balpha_{1} \cdot \balpha_{2}}{r_{12}} 
    + \frac{1}{2}\left[\balpha_{1} \cdot \bnabla_{1},\left[\balpha_{2} \cdot \bnabla_{2},r_{12}\right] \right]
  \right)\,,
\label{eq:breit0} 
\end{equation}
with the summation over intermediate states restricted to the positive-energy spectrum.

\begin{table}
\caption{\label{T1_IIC}Interelectronic-interaction contributions to the $g$ factor of $(1s)^2 2p_{1/2}$ state of lithium-like ions obtained in the Coulomb and different screening potentials: core-Hartree (CH), Kohn-Sham (KS), Dirac-Hartree (DH), and Dirac-Slater (DS), in units of $10^{-6}$.}
\begin{center}
\begin{tabular}{c S S S S S}
\hline\hline
 &\multicolumn{1}{c}{Coulomb} &\multicolumn{1}{c}{CH} &\multicolumn{1}{c}{KS} &\multicolumn{1}{c}{DH} &\multicolumn{1}{c}{DS}\\
\hline
\multicolumn{6}{l}{$Z = 14$}\\
$\Delta g^{(0)}_{\rm int}$ 
                    &                & 427.0705    & 402.3588   & 506.2160    & 348.2924     \\
$\dgint^{(1)}$      & 428.1075 (1)       & -20.1230 (5)   & 6.2167 (2)     &-102.9122 (4)   &  64.1068 (1)     \\
$\Delta g_{\rm int, L}^{(2)}$        & -19.5850 (1)    & 1.5841     &  -0.2063    &   5.3909 (1)  &  -4.2914  \\     
$\Delta g_{\rm int, H}^{(2)}$        & \pm 0.204 \\
$\Delta g_{\rm int, H}^{(3+)}$       & \pm 1.399 \\
Total               & 408.5 (1.4)   & 408.5 (14) &  408.4 (14)  & 408.7 (14) &  408.1 (14) \\
\hline                                                        
\multicolumn{6}{l}{$Z = 54$}\\
$\Delta g^{(0)}_{\rm int}$ 
                    &                & 1826.1207     & 1691.3712     & 2188.5094      & 1440.2618      \\
$\dgint^{(1)}$      & 1769.6284      & -78.4565 (17)    & 59.7164 (6)    & -444.1441 (14)      &  315.5818 (1)      \\
$\Delta g_{\rm int, L}^{(2)}$      &   -23.7582   &  1.1515  &  -2.7774 (1)  &   4.9831 (1)   &   -8.0934 (1)  \\
$\Delta g_{\rm int, H}^{(2)}$        & \pm 3.689 \\
$\Delta g_{\rm int, H}^{(3+)}$       & \pm 0.440 \\
Total               &  1745.9 (37)   &  1748.8 (37)  &  1748.3 (37)  &  1749.3 (37)     &  1747.8 (37) \\
\hline
\multicolumn{6}{l}{$Z = 82$}\\
$\Delta g^{(0)}_{\rm int}$ 
                    &                & 3099.2479     & 2849.2842     & 3724.1368      & 2408.6054      \\
$\dgint^{(1)}$      & 3033.8546 (10)      & -90.0362 (16)    & 166.5854 (1)    & -717.8554 (12)     &  613.8629 (8)     \\
$\Delta g_{\rm int, L}^{(2)}$      &   -28.7508 (3)  &  0.1643 (2) &  -7.2666 (3)  &   3.9237 (2)   &   -14.7004 (3)  \\
$\Delta g_{\rm int, H}^{(2)}$        & \pm 10.295 \\
$\Delta g_{\rm int, H}^{(3+)}$       & \pm 0.351 \\
Total               &  3005.1 (103)   &  3009.4 (103) &  3008.6 (103)   &  3010.2 (103)    &  3007.8 (103) \\  
\hline\hline
\end{tabular}
\end{center}
\end{table}

\begin{table}
\caption{\label{T2_IIC}Interelectronic-interaction contributions to the $g$ factor of $(1s)^2 2p_{3/2}$ state of lithium-like ions obtained in the Coulomb and different screening potentials: core-Hartree (CH), Kohn-Sham (KS), Dirac-Hartree (DH), and Dirac-Slater (DS), in units of $10^{-6}$.}
\begin{center}
\begin{tabular}{c S S S S S}
\hline\hline
 &\multicolumn{1}{c}{Coulomb} &\multicolumn{1}{c}{CH} &\multicolumn{1}{c}{KS} &\multicolumn{1}{c}{DH} &\multicolumn{1}{c}{DS}\\
\hline
\multicolumn{6}{l}{$Z = 14$}\\
$\Delta g^{(0)}_{\rm int}$ 
                    &                & 340.6594    & 321.1211        & 403.7611       & 278.1117     \\
$\dgint^{(1)}$      & 348.1671 (3)      & -8.3504 (1)   & 12.3569 (1)        & -74.3469 (1)      &  58.3216 (2)    \\
$\Delta g_{\rm int, L}^{(2)}$      &  -14.9848 (3)   &   0.6683 (1) &  -0.6177 (1)    &   3.7328    &  -3.7776 (1) \\
$\Delta g_{\rm int, H}^{(2)}$        & \pm 0.156 \\
$\Delta g_{\rm int, H}^{(3+)}$       & \pm 1.070 \\
Total               & 333.2 (11)  & 333.0 (11) &  332.9 (11)  & 333.1 (11) &  332.7 (11) \\
\hline                    
\multicolumn{6}{l}{$Z = 54$}\\
$\Delta g^{(0)}_{\rm int}$ 
                    &                & 1395.8880     & 1304.4037     & 1669.6127      & 1120.1069      \\
$\dgint^{(1)}$      & 1364.8419 (13)     & -46.8100 (2)    & 46.2699 (7)    & -322.9978      &  233.5202 (10)      \\
$\Delta g_{\rm int, L}^{(2)}$      &   -18.7276 (4)   &  0.4415 (2) &  -1.5608 (2)  &   3.5079 (1)   &   -5.0520 (2)  \\
$\Delta g_{\rm int, H}^{(2)}$        & \pm 2.908 \\
$\Delta g_{\rm int, H}^{(3+)}$       & \pm 0.347 \\
Total               &  1346.1 (29)   &  1349.5 (29)  &  1349.1 (29)  &  1350.1 (29)     &  1348.6 (29) \\
\hline
\multicolumn{6}{l}{$Z = 82$}\\
$\Delta g^{(0)}_{\rm int}$ 
                    &                & 2203.9020     & 2070.4084     & 2633.3109      & 1787.3028      \\
$\dgint^{(1)}$      & 2143.8537 (21)     & -77.8406 (3)    & 57.7889 (11)    & -509.4014      &  343.9877 (16)     \\
$\Delta g_{\rm int, L}^{(2)}$      &   -24.5899 (6)  &  0.0293 (3) &  -2.8107 (3) &   3.3202 (2)   &   -6.8530 (4)  \\
$\Delta g_{\rm int, H}^{(2)}$        & \pm 8.805 \\
$\Delta g_{\rm int, H}^{(3+)}$       & \pm 0.300 \\
Total               &  2119.3 (88)  &  2126.1 (88)  &  2125.4 (88)  &  2127.2 (88)    &  2124.4 (88) \\  
\hline\hline
\end{tabular}
\end{center}
\end{table}

%
\subsection{QED corrections}
\label{sec:qed}

One-loop QED corrections to the $g$ factor consist of self-energy and vacuum polarization contributions:
\begin{equation}
  \Delta g_{\rm QED} = \Delta g_{\rm SE} + \Delta g_{\rm VP}
\,.
\end{equation}
The self-energy correction for the $2p_j$ states has been calculated to all orders in $\alpha Z$ for $Z=1-12$ in \cite{Yerokhin:2010:012502}.
In this work, we employ the method developed in \cite{Glazov:2006:330, Volotka:2006:293} which is based on a finite basis set constructed from B-splines \cite{Shabaev:2004:130405}. 
To evaluate the self-energy correction, it is necessary to eliminate the ultraviolet divergence. For this purpose, the 0-potential and 1-potential contributions are separated as in \cite{Yerokhin:2004:052503}. The ultraviolet finite part is then calculated in momentum space, while the remaining many-potential term is computed in coordinate space using a multipole expansion. The contributions are divided by the angular quantum number $\kappa$, and for each contribution, a double integration over the coordinates is performed, followed by an integration over $\omega$ in the complex plane, using the contour described in \cite{Yerokhin:1999:800}.

The vacuum polarization correction includes the electric-loop $\dgvpel$, and the magnetic-loop $\dgvpm$ contributions. The electric-loop part is treated within the Uehling approximation. For the magnetic-loop contribution, we consider terms up to second order in $\alpha Z$, using the results from \cite{Lee:07:CJP}. The uncertainty from the vacuum polarization correction is given by the root-mean-square of two terms, namely the Wichman-Kroll uncertainty and the magnetic-loop uncertainty. The Wichman-Kroll uncertainty is estimated by extrapolating the Wichman-Kroll correction to the $g$ factor of lithium-like ions from~\cite{Glazov:2004:062104}, yielding an estimate of $0.2(\alpha Z)^{2} \dgvpel$. The uncertainty from the magnetic loop we estimate as $\alpha Z$ multiplied by the lowest calculated order of $\alpha Z$. 

To approximate many-electron QED effects, we use effective screening potentials consistent with those applied for interelectronic-interaction contributions. Such calculations for the ground state of boron-like ions have been performed in \cite{Agababaev:2018:012003}.

The two-loop QED contributions are evaluated within the $\alpha Z$-expansion framework. The most recent results for $s$ states are available in \cite{Pachucki:2005:022108, Czarnecki:2020:050801}, and for $p$ states in \cite{Jentschura:2010:012512}.

\begin{table}
\caption{\label{T3_QED}QED contributions to the $g$ factor of $(1s)^2 2p_{1/2}$ state of lithium-like ions obtained in the Coulomb and different screening potentials: core-Hartree (CH), Kohn-Sham (KS), Dirac-Hartree (DH), and Dirac-Slater (DS), in units of $10^{-6}$.}
\begin{center}
\resizebox{0.7\textwidth}{!}{%
\begin{tabular}{c S S S S S}
\hline\hline
 &\multicolumn{1}{c}{Coulomb} &\multicolumn{1}{c}{CH} &\multicolumn{1}{c}{KS} &\multicolumn{1}{c}{DH} &\multicolumn{1}{c}{DS}\\
\hline
\multicolumn{6}{l}{$Z = 14$}\\
$\dgirrlo$& -9.615& -9.054& -11.004& -10.525& -11.246\\
$\dgirrfo$& -29.761& -22.736& -21.850& -20.260& -22.670\\
$\dgirrm$& -0.021& 0.772& 1.490& 1.447& 1.508\\
$\dgvrlo$& -797.529& -792.610& -792.447& -791.330& -793.019\\
$\dgvrfo$& 46.925& 37.761& 37.982& 35.725& 39.137\\
$\dgvrm$& 19.096& 14.383& 14.444& 13.371& 15.002\\
$\dgvpel$& -0.000& -0.000& -0.000& -0.000& -0.000\\
$\dgvpm$& 0.000& 0.000& 0.000& 0.000& 0.000\\
$\dgqed$& -770.90(8)& -771.49(11)& -771.39(11)& -771.57(11)& -771.29(11)\\
\hline                    
\multicolumn{6}{l}{$Z = 54$}\\
$\dgirrlo$& -26.755& -29.320& -32.957& -32.954& -32.956\\
$\dgirrfo$& -236.930& -226.165& -225.305& -222.503& -226.706\\
$\dgirrm$& 24.019& 25.327& 28.227& 27.797& 28.439\\
$\dgvrlo$& -949.079& -941.615& -941.349& -939.500& -942.273\\
$\dgvrfo$& 282.667& 272.051& 272.439& 269.555& 273.883\\
$\dgvrm$& 222.815& 210.443& 210.593&  207.432& 212.181\\
$\dgvpel$& -0.763& -0.683& -0.702& -0.677& -0.715\\
$\dgvpm$& 0.110(6)& 0.110(6)&  0.110(6)& 0.110(6)&  0.110(6)\\
$\dgqed$& -683.53(8)& -689.5(10)& -688.6(10)& -690.4(10)& -687.7(10)\\
\hline
\multicolumn{6}{l}{$Z = 82$}\\
$\dgirrlo$& -41.515& -43.945& -48.310& -48.010& -48.462\\
$\dgirrfo$& -328.048& -324.246& -322.617& -321.804& -323.006\\
$\dgirrm$& 99.514& 99.297& 103.463& 102.258& 104.062\\
$\dgvrlo$& -1046.379& -1041.268& -1040.739& -1039.532& -1041.336\\
$\dgvrfo$& 436.481& 427.331& 427.617& 425.079& 428.885\\
$\dgvrm$& 405.297& 393.883& 393.900& 390.922& 395.388\\
$\dgvpel$& -7.251& -6.684& -6.830& -6.642& -6.926\\
$\dgvpm$& 0.95(12)& 0.95(12)&  0.95(12)& 0.95(12)&  0.95(12)\\
$\dgqed$& -481.0(2.0)& -494.7(2.4)& -492.6(2.4)& -496.8(2.4)& -490.4(2.4)\\  
\hline\hline
\end{tabular}}
\end{center}
\end{table} 

\begin{table}
\caption{\label{T4_QED}QED contributions to the $g$ factor of $(1s)^2 2p_{3/2}$ state of lithium-like ions obtained in the Coulomb and different screening potentials: core-Hartree (CH), Kohn-Sham (KS), Dirac-Hartree (DH), and Dirac-Slater (DS), in units of $10^{-6}$.}
\begin{center}
\resizebox{0.7\textwidth}{!}{%
\begin{tabular}{c S S S S S}
\hline\hline
 &\multicolumn{1}{c}{Coulomb} &\multicolumn{1}{c}{CH} &\multicolumn{1}{c}{KS} &\multicolumn{1}{c}{DH} &\multicolumn{1}{c}{DS}\\
\hline
\multicolumn{6}{l}{$Z = 14$}\\
$\dgirrlo$& 19.761& 14.334& 12.434& 11.520& 12.905\\
$\dgirrfo$& -3.878& -1.899& -0.141& -0.028& -0.200\\
$\dgirrm$& 3.591& 2.930& 3.289& 3.067& 3.404\\
$\dgvrlo$& 736.888& 744.598& 744.978& 746.707& 744.093\\
$\dgvrfo$& 13.413& 10.804& 10.138& 9.683& 10.367\\
$\dgvrm$& 8.268& 6.474& 6.623& 6.155& 6.871\\
$\dgvp$& -0.000& -0.000& -0.000& -0.000& -0.000\\
$\dgqed$& 778.04(10)& 777.24(12)& 777.32(12)& 777.11(12)& 777.44(12)\\
\hline                    
\multicolumn{6}{l}{$Z = 54$}\\
$\dgirrlo$& 201.091& 188.659& 185.260& 182.715& 186.537\\
$\dgirrfo$& -71.654& -66.412& -63.772& -62.934& -64.192\\
$\dgirrm$& 36.102& 35.002& 36.263& 35.782& 36.437\\
$\dgvrlo$& 510.176& 521.844& 522.339& 525.137& 520.938\\
$\dgvrfo$& 53.147& 51.556& 50.321& 50.223& 50.367\\
$\dgvrm$& 116.327& 109.802& 110.557& 108.745& 111.469\\
$\dgvp$& -0.051& -0.045& -0.046& -0.045& -0.047\\
$\dgvpm$& 0.188& 0.188& 0.188& 0.188& 0.188\\
$\dgqed$& 845.33(50)& 840.59(70)& 841.11(70)& 839.81(70)& 841.70(70)\\
\hline
\multicolumn{6}{l}{$Z = 82$}\\
$\dgirrlo$& 370.780& 357.817& 354.212& 351.549& 355.544\\
$\dgirrfo$& -132.086& -127.725& -125.219& -124.539& -125.556\\
$\dgirrm$& 67.112& 65.937& 67.745& 67.185& 68.075\\
$\dgvrlo$& 338.833& 349.996& 350.512& 353.134& 349.201\\
$\dgvrfo$& 63.033& 62.081& 60.707& 60.793& 60.663\\
$\dgvrm$& 246.531& 237.722& 238.718& 236.375& 239.893\\
$\dgvpel$& -0.553& -0.507& -0.516& -0.503& -0.522\\
$\dgvpm$& 1.520(5)& 1.520(5)& 1.520(5)& 1.520(5)&  1.520(5)\\
$\dgqed$& 955.2(1.0)& 946.8(1.2)& 947.7(1.2)& 945.5(1.2)& 948.8(1.2)\\ 
\hline\hline
\end{tabular}}
\end{center}
\end{table}

\subsection{Nuclear-recoil effects}
\label{sec:rec}

The first fully relativistic treatment of the nuclear recoil correction to the $g$ factor was developed in~\cite{Shabaev:2001:052104}, where the QED formalism was extended beyond the traditional Furry picture. Subsequent studies have applied this formalism to hydrogen-like ions across a wide range of nuclear charges, including the $1s$, $2s$, $2p_{1/2}$, and $2p_{3/2}$ states~\cite{Malyshev:2020:012513}. These calculations account for all orders in $\alpha Z$ and are accurate to first order in the small parameter $m/M$, where $m$ is the electron mass and $M$ is the nuclear mass.

For lithium-like ions, which possess additional interelectronic correlations, the recoil contribution has been treated using effective four-component recoil operators within the Breit approximation~\cite{Shabaev:2017:263001}. This method has enabled high-precision theoretical predictions for recoil corrections in few-electron ions, which are in strong demand due to current and upcoming experiments, such as those at the ARTEMIS facility at GSI and the ALPHATRAP experiment at MPIK~\cite{Glazov:2019:173001, Arapoglou:2019:253001}.

In the following, we present the evaluation of the nuclear recoil correction to the bound-electron $g$ factor. The calculation is performed to first order in $m/M$, and includes both the nonmagnetic and magnetic contributions derived within the Breit approximation. These terms incorporate the leading relativistic recoil effects and provide a reliable basis for interpreting high-precision spectroscopic data.

To first order in the electron-to-nucleus mass ratio $m/M$, this effect is described by effective recoil operators derived in \cite{Shabaev:2017:263001}:

\begin{equation}
  \hml = \frac{1}{2M}\, \sum_{j,k}
    \left[ \bfp_j\cdot \bfp_k - \frac{\alpha Z}{r_j}
      \left( \balpha_j + \frac{(\balpha_j\cdot\bfr_j)\bfr_j}{r_j^2} \right)
      \cdot \bfp_k \right] \,,
\end{equation}

\begin{equation}
  \hmm = -\mu_0 {\bf B} \frac{m}{M} \sum_{j,k}\bigg\{
    [\bfr_j \times  \bfp_k] - \frac{\aZ}{2r_k}
    \left[\bfr_j \times
    \left( \balpha_k
          + \frac{(\balpha_k\cdot\bfr_k)\bfr_k}{r_k^2}
        \right)\right]\bigg\} \,.
\end{equation} 

These operators account for all contributions up to order $(m/M)(\alpha Z)^2$.
The $g$-factor recoil correction associated with $\hml$ is calculated using second-order perturbation theory:
\begin{equation}
\label{gnonmagn}
    \dgrecnonmgn =  \frac{2}{M_J} \sum_{N\neq A} \frac{\matrixel{A}{\hml}{N}\matrixel{N}{U}{A}}{E_a - E_N} \,.
\end{equation}

The correction associated with $H_{\rm M}^{\rm magn}$ is calculated using first-order perturbation theory:

\begin{equation}
\label{gmagn}
    \dgrecmgn =  \frac{1}{\mu_0 B M_J} \matrixel{A}{\hmm}{A} \,.
\end{equation}


The total nuclear recoil correction is given by
\begin{equation}
    \dgrecfull = \dgrecnonmgn + \dgrecmgn \,.
\end{equation}





%
\section{RESULTS AND DISCUSSIONS}
\label{sec:results}

In this section, we discuss a numerical evaluation of all considered contributions and present the results for the excited states of lithium-like ions. Following our previous work~\cite{Zinenko:2023:032815}, all calculations are based on the dual-kinetically-balanced finite-basis-set method~\cite{Shabaev:2004:130405} for the Dirac equation with basis functions constructed from B-splines~\cite{Sapirstein:1996:5213}.

We first solve the Dirac equation~\eqref{eq:Dirac} with one of the chosen effective potentials. The zeroth-order interelectronic-interaction contribution $\Delta g^{(0)}_{\mathrm{int}}$ is then found according to Eq.~\eqref{eq:g_scr}. In this work, we calculate it using various screening potentials with the required accuracy.

The second order contribution, $\Delta g^{(2)}_{\mathrm{int}}$, is calculated within the Breit approximation. To ensure convergence, the number of basis functions is increased up to $N = 120$, followed by an extrapolation to the limit $N \to \infty$. The partial-wave summation over the relativistic angular quantum number $\kappa=(j+1/2)^{j+l+1/2}$ is truncated at $|\kappa_{\rm max}| = 10$, and the remaining contribution is estimated by an inverse polynomial least-squares fit.

Table~\ref{T1_IIC} presents the interelectronic-interaction contributions $\dgint$ to the $g$ factor of the $(1s)^2\,2p_{1/2}$ state in lithium-like ions for $Z = 14, 54, 82$. The results are obtained with the Coulomb and different screening potentials: core-Hartree (CH), Kohn-Sham (KS), Dirac-Hartree (DH), and Dirac-Slater (DS). Since the second-order contribution is known only in the leading approximation, denoted $\Delta g_{\rm int, L}^{(2)}$, the unknown higher-order part $\Delta g_{\rm int, H}^{(2)}$ is estimated as $\Delta g_{\rm int, L}^{(2)} (\alpha Z)^2$ within the Coulomb potential. The third and higher orders are still unknown, so we estimate them by a simple expression $\Delta g_{\rm int, H}^{(3+)} \simeq \Delta g_{\rm int, L}^{(2)}/Z$. As seen from the Table, the total values obtained using different screening potentials are quite close to each other and overlap within their uncertainties. Table~\ref{T2_IIC} shows similar results for the $(1s)^2\,2p_{3/2}$ state.

Tables~\ref{T3_QED} and \ref{T4_QED} show the results of the QED corrections to the $g$ factor of lithium-like ions for the $(1s)^2\,2p_{1/2}$ and $(1s)^2\,2p_{3/2}$ states, respectively, for $Z = 14, 54, 82$. At low $Z$, the vacuum polarization contribution $\dgvp$ is negligible compared to the total uncertainty. However, with increasing $Z$, this contribution becomes significant and must be taken into account in the final result.

We also account for the leading nonlinear Zeeman effect contributions to the $g$ factor. From the standpoint of experimental Zeeman splitting measurements, these corrections can be interpreted as effective modifications to the $g$ factor. In the presence of a magnetic field, the energy of a state $|a\rangle$ is modified as follows:

\begin{equation}
\label{deltag}
    E = E^{(0)} + \mub B M_a (g + \mub B \delta^{(2)} g  + (\mub  B)^2 \delta^{(3)} g  + \dots),
\end{equation}

In the absence of a magnetic field, the fine structure splitting between levels with distinct total angular momentum, represented by $j_a$, is observed. A magnetic field lifts the degeneracy with respect to the magnetic quantum number $M_a$. To first order in magnetic field $B$, the energy shift is linear in $M_a$, with the $g$ factor as the proportionality constant. The second-order shift is independent of the sign of $M_a$ and hence leads to a $g$ factor correction $\delta^{(2)} g_a$ that depends on the magnitude but not the sign of $M_a$. The corresponding results for the quadratic effect correction as a modification to the $g$ factor are given in Table~\ref{T5_G2}. The third-order shift depends on the sign of the projection $M_a$, similar to the linear term, and its contribution to the $g$ factor is summarized in Table~\ref{T6_G3}.
 

Table~\ref{tab:g_2p1} summarizes the complete theoretical results for the $g$ factor of the excited states of lithium-like ions over a wide range of nuclear charge numbers, $Z=10-92$. The CH potential is used for these values, as it is uniquely defined, in contrast to other $x_\alpha$ potentials. 

\begin{table}
\caption{\label{T5_G2}Quadratic Zeeman effect represented as the $g$-factor correction $\delta^{(2)} g$ at the field of 1 T (see Eq. \eqref{deltag}) of $(1s)^2 2p_{1/2}$ and $(1s)^2 2p_{3/2}$ states of lithium-like ions. The value of $g^{(2)}$ for the core-Hartree potential is used.}
\begin{center}
\begin{tabular}{c S S S}
\hline\hline
 \multicolumn{1}{c}{Z} & \multicolumn{1}{c}{$2p_{1/2}$} & \multicolumn{2}{c}{$2p_{3/2}$} \\
 & \multicolumn{1}{c}{$M_J = \pm 1/2$} & \multicolumn{1}{r}{$M_J = \pm 1/2$} &  \multicolumn{1}{r}{$M_J = \pm 3/2$} \\
\hline
 10 & \mp 1.1   $\times 10^{-4}$  & \pm 1.2   $\times 10^{-4}$  & \pm2.6 $\times 10^{-7}$ \\
 12 & \mp4.8   $\times 10^{-5}$  & \pm4.9   $\times 10^{-5}$  & \pm1.7 $\times 10^{-7}$ \\
 14 & \mp2.4   $\times 10^{-5}$  & \pm2.4   $\times 10^{-5}$  & \pm1.2 $\times 10^{-7}$ \\
 16 & \mp1.3   $\times 10^{-5}$  & \pm1.3   $\times 10^{-5}$  & \pm8.5 $\times 10^{-8}$ \\
 18 & \mp7.6   $\times 10^{-6}$  & \pm7.9   $\times 10^{-6}$  & \pm6.5 $\times 10^{-8}$ \\
 20 & \mp4.8   $\times 10^{-6}$  & \pm5.0   $\times 10^{-6}$  & \pm5.1 $\times 10^{-8}$ \\
 24 & \mp2.1   $\times 10^{-6}$  & \pm2.3   $\times 10^{-6}$  & \pm3.4 $\times 10^{-8}$ \\
 32 & \mp5.9   $\times 10^{-7}$  & \pm6.8   $\times 10^{-7}$  & \pm1.8 $\times 10^{-8}$ \\
 40 & \mp2.2   $\times 10^{-7}$  & \pm2.7   $\times 10^{-7}$  & \pm1.1 $\times 10^{-8}$ \\
 54 & \mp5.3   $\times 10^{-8}$  & \pm7.8   $\times 10^{-8}$  & \pm5.9 $\times 10^{-9}$  \\
 70 & \mp1.4   $\times 10^{-8}$  & \pm2.7   $\times 10^{-8}$  & \pm3.3 $\times 10^{-9}$  \\
 82 & \mp5.1   $\times 10^{-9}$  & \pm1.5   $\times 10^{-8}$  & \pm2.3 $\times 10^{-9}$  \\
 92 & \mp2.2   $\times 10^{-9}$  & \pm9.1   $\times 10^{-10}$ & \pm1.7 $\times 10^{-9}$  \\
\hline\hline
\end{tabular}
\end{center}
\end{table}

\begin{table}
\caption{\label{T6_G3}Cubic Zeeman effect represented as the $g$-factor correction $\delta^{(3)} g$ at the field of 1 T (see Eq. \eqref{deltag}) of $(1s)^2 2p_{1/2}$ and $(1s)^2 2p_{3/2}$ states of lithium-like ions. The value of $g^{(3)}$ for the core-Hartree potential is used.}
\begin{center}
\begin{tabular}{c S S S}
\hline\hline
 \multicolumn{1}{c}{Z} & \multicolumn{1}{c}{$2p_{1/2}$} & \multicolumn{2}{c}{$2p_{3/2}$} \\
 & \multicolumn{1}{c}{$M_J = \pm 1/2$} & \multicolumn{1}{r}{$M_J = \pm 1/2$} &  \multicolumn{1}{r}{$M_J = \pm 3/2$} \\
\hline
 10 & 1.0   $\times 10^{-8}$  & -1.0   $\times 10^{-8}$  & -1.7 $\times 10^{-16}$ \\
 12 & 1.8   $\times 10^{-9}$  & -1.8   $\times 10^{-9}$  & -1.1 $\times 10^{-16}$ \\
 14 & 4.4   $\times 10^{-10}$ & -4.4   $\times 10^{-10}$ & -7.5 $\times 10^{-17}$ \\
 16 & 1.3   $\times 10^{-10}$ & -1.3   $\times 10^{-10}$ & -5.5 $\times 10^{-17}$ \\
 18 & 4.6   $\times 10^{-11}$ & -4.6   $\times 10^{-11}$ & -4.2 $\times 10^{-17}$ \\
 20 & 1.8   $\times 10^{-11}$ & -1.8   $\times 10^{-11}$ & -3.3 $\times 10^{-17}$ \\
 24 & 3.8   $\times 10^{-12}$ & -3.8   $\times 10^{-12}$ & -2.2 $\times 10^{-17}$ \\
 32 & 3.3   $\times 10^{-13}$ & -3.3   $\times 10^{-13}$ & -1.2 $\times 10^{-17}$ \\
 40 & 5.0   $\times 10^{-14}$ & -5.0   $\times 10^{-14}$ & -7.3 $\times 10^{-18}$ \\
 54 & 4.0   $\times 10^{-15}$ & -4.0   $\times 10^{-15}$ & -3.8 $\times 10^{-18}$  \\
 70 & 4.5   $\times 10^{-16}$ & -4.5   $\times 10^{-16}$ & -2.2 $\times 10^{-18}$  \\
 82 & 1.1   $\times 10^{-16}$ & -1.1   $\times 10^{-16}$ & -1.5 $\times 10^{-18}$  \\
 92 & 4.0   $\times 10^{-17}$ & -4.0   $\times 10^{-17}$ & -1.2 $\times 10^{-18}$  \\
\hline\hline
\end{tabular}
\end{center}
\end{table}

\begin{table}
\caption{\label{tab:g_2p1}
Individual contributions to the $(1s)^2 2p_{1/2}$ and $(1s)^2 2p_{3/2}$  states of $g$ factor in lithium-like ions in the range $Z=10$--$92$. The CH potential was chosen for the obtained corrections $\dgint$, $\dgqed$ and $\dgnuc$.}
\vspace{0.5cm}
\begin{center}
\begin{tabular}{lr@{}lr@{}l}
${}^{20}_{10}$Ne$^{7+}$ \\
\hline 
& \multicolumn{2}{c}{$(1s)^2 2p_{1/2}$}
& \multicolumn{2}{c}{$(1s)^2 2p_{3/2}$}
\\[6pt]
\hline 
Dirac value $\gdirac$                     &    0.&665\,777\,663     &    1.&332\,623\,079     \\
Interelectronic interaction $\dgint$  &    0.&000\,285\,5\,(19) &    0.&000\,233\,2\,(15) \\
One-loop QED $\dgqed$               & $-$0.&000\,773\,01\,(6)  & 0.&000\,775\,62\,(8)  \\
Two-loop QED $\dgqed$               &    0.&000\,001\,2\,(1)       &    0.&000\,001\,2\,(1)       \\
Nuclear recoil $\dgnuc$               & $-$0.&000\,024\,4\,(21) & $-$0.&000\,012\,2\,(11)  \\
\hline 
Total value $g$                       &    0.&665\,267\,0\,(28) &    1.&333\,620\,9\,(19) \\
\hline
\\
${}^{24}_{12}$Mg$^{9+}$ \\
\hline
& \multicolumn{2}{c}{$(1s)^2 2p_{1/2}$}
& \multicolumn{2}{c}{$(1s)^2 2p_{3/2}$}
\\[6pt]
\hline 
Dirac value $\gdirac$                     &    0.&665\,385\,559     &    1.&332\,310\,417     \\
Interelectronic interaction $\dgint$  &    0.&000\,346\,9\,(16) &    0.&000\,283\,1\,(12)  \\
One-loop QED $\dgqed$               & $-$0.&000\,772\,34\,(9)  & 0.&000\,776\,35\,(10)  \\
Two-loop QED $\dgqed$               &    0.&000\,001\,2\,(1)       &    0.&000\,001\,2\,(1)  \\
Nuclear recoil $\dgnuc$               & $-$0.&000\,019\,9\,(14)  & $-$0.&000\,009\,98\,(75)  \\
\hline 
Total value $g$                       &    0.&664\,941\,4\,(21) &    1.&333\,361\,1\,(14) \\
\hline
\\
${}^{28}_{14}$Si$^{11+}$ \\
\hline
& \multicolumn{2}{c}{$(1s)^2 2p_{1/2}$}
& \multicolumn{2}{c}{$(1s)^2 2p_{3/2}$}
\\[6pt]
\hline 
Dirac value $\gdirac$                     &    0.&664\,921\,417     &    1.&331\,940\,789     \\
Interelectronic interaction $\dgint$  &    0.&000\,408\,5\,(14)  &    0.&000\,333\,0\,(11)  \\
One-loop QED $\dgqed$               & $-$0.&000\,771\,49\,(11)  & 0.&000\,777\,24\,(12) \\
Two-loop QED $\dgqed$               &    0.&000\,001\,2\,(1)  &    0.&000\,001\,2\,(1)  \\
Nuclear recoil $\dgnuc$               & $-$0.&000\,016\,8\,(11)  & $-$0.&000\,008\,42\,(55)  \\
\hline 
Total value $g$                       &    0.&664\,542\,8\,(18) &    1.&333\,043\,8\,(12) \\
\hline
\end{tabular}
\end{center}
\end{table}

\begin{table}
\begin{center}
\begin{tabular}{lr@{}lr@{}l}
${}^{32}_{16}$S$^{13+}$ \\
\hline
& \multicolumn{2}{c}{$(1s)^2 2p_{1/2}$}
& \multicolumn{2}{c}{$(1s)^2 2p_{3/2}$}
\\[6pt]
\hline 
Dirac value $\gdirac$                     &    0.&664\,384\,860     &    1.&331\,514\,136     \\
Interelectronic interaction $\dgint$  &    0.&000\,470\,4\,(13)  &    0.&000\,382\,9\,(10)  \\
One-loop QED $\dgqed$               & $-$0.&000\,770\,45\,(13)  & 0.&000\,778\,31\,(14) \\
Two-loop QED $\dgqed$               &    0.&000\,001\,2\,(1)  &    0.&000\,001\,2\,(1)  \\
Nuclear recoil $\dgnuc$               & $-$0.&000\,014\,51\,(84)  & $-$0.&000\,007\,29\,(42)  \\
\hline 
Total value $g$                       &    0.&664\,071\,5\,(16) &    1.&332\,669\,3\,(11) \\
\hline
\\
${}^{40}_{18}$Ar$^{15+}$ \\
\hline
& \multicolumn{2}{c}{$(1s)^2 2p_{1/2}$}
& \multicolumn{2}{c}{$(1s)^2 2p_{3/2}$}
\\[6pt]
\hline  
Dirac value $\gdirac$                     &    0.&663\,775\,447     &    1.&331\,030\,389     \\
Interelectronic interaction $\dgint$  &    0.&000\,532\,5\,(12)  &    0.&000\,432\,8\,(9)  \\
One-loop QED $\dgqed$               & $-$0.&000\,769\,20\,(15)  & 0.&000\,779\,57\,(17) \\
Two-loop QED $\dgqed$               &    0.&000\,001\,2\,(1)  &    0.&000\,001\,2\,(1)  \\
Nuclear recoil $\dgnuc$               & $-$0.&000\,011\,50\,(60)  & $-$0.&000\,005\,78\,(30)  \\
\hline 
Total value $g$                       &    0.&663\,528\,5\,(14) &    1.&332\,238\,2\,(10) \\
\hline
\\
${}^{40}_{20}$Ca$^{17+}$ \\
\hline
& \multicolumn{2}{c}{$(1s)^2 2p_{1/2}$}
& \multicolumn{2}{c}{$(1s)^2 2p_{3/2}$}
\\[6pt]
\hline 
Dirac value $\gdirac$                     &    0.&663\,092\,678     &    1.&330\,489\,471     \\
Interelectronic interaction $\dgint$  &    0.&000\,594\,9\,(11)  &    0.&000\,482\,8\,(8)  \\
One-loop QED $\dgqed$               & $-$0.&000\,767\,74\,(18)  & 0.&000\,781\,07\,(19) \\
Two-loop QED $\dgqed$               &    0.&000\,001\,2\,(1)  &    0.&000\,001\,2\,(1)  \\
Nuclear recoil $\dgnuc$               & $-$0.&000\,011\,40\,(54)  & $-$0.&000\,005\,74\,(27)  \\
\hline 
Total value $g$                       &    0.&662\,909\,6\,(12) &    1.&331\,748\,8\,(9) \\
\hline \hline
\end{tabular}
\end{center}
\end{table}

\begin{table}
\begin{center}
\begin{tabular}{lr@{}lr@{}l}
${}^{52}_{24}$Cr$^{21+}$ \\
\hline  
& \multicolumn{2}{c}{$(1s)^2 2p_{1/2}$}
& \multicolumn{2}{c}{$(1s)^2 2p_{3/2}$}
\\[6pt]
\hline 
Dirac value $\gdirac$                     &    0.&661\,504\,731     &    1.&329\,235\,759     \\
Interelectronic interaction $\dgint$  &    0.&000\,720\,8\,(10)  &    0.&000\,583\,0\,(8)  \\
One-loop QED $\dgqed$               & $-$0.&000\,764\,07\,(24)  & 0.&000\,784\,48\,(24) \\
Two-loop QED $\dgqed$               &    0.&000\,001\,2\,(1)  &    0.&000\,001\,2\,(1)  \\
Nuclear recoil $\dgnuc$               & $-$0.&000\,008\,67\,(34)  & $-$0.&000\,004\,38\,(17)  \\
\hline 
Total value $g$                       &    0.&661\,454\,0\,(11) &    1.&330\,600\,1\,(9) \\
\hline
\\
${}^{74}_{32}$Ge$^{29+}$ \\
\hline 
& \multicolumn{2}{c}{$(1s)^2 2p_{1/2}$}
& \multicolumn{2}{c}{$(1s)^2 2p_{3/2}$}
\\[6pt]
\hline 
Dirac value $\gdirac$                     &    0.&657\,418\,976     &    1.&326\,037\,799     \\
Interelectronic interaction $\dgint$  &    0.&000\,978\,2\,(13)  &    0.&000\,784\,2\,(10)  \\
One-loop QED $\dgqed$               & $-$0.&000\,753\,25\,(39)  & 0.&000\,793\,97\,(36) \\
Two-loop QED $\dgqed$               &    0.&000\,001\,2\,(1)  &    0.&000\,001\,2\,(1)  \\
Nuclear recoil $\dgnuc$               & $-$0.&000\,005\,98\,(18)  & $-$0.&000\,003\,05\,(9)  \\
\hline 
Total value $g$                       &    0.&657\,639\,1\,(14) &    1.&327\,614\,1\,(11) \\
\hline
\\
${}^{91}_{40}$Zr$^{37+}$ \\
\hline 
& \multicolumn{2}{c}{$(1s)^2 2p_{1/2}$}
& \multicolumn{2}{c}{$(1s)^2 2p_{3/2}$}
\\[6pt]
\hline 
Dirac value $\gdirac$                     &    0.&652\,070\,328     &    1.&321\,911\,896     \\
Interelectronic interaction $\dgint$  &    0.&001\,245\,5\,(19)  &    0.&000\,987\,3\,(15)  \\
One-loop QED $\dgqed$               & $-$0.&000\,736\,67\,(57)  & 0.&000\,807\,21\,(48) \\
Two-loop QED $\dgqed$               &    0.&000\,001\,2\,(1)  &    0.&000\,001\,2\,(1)  \\
Nuclear recoil $\dgnuc$               & $-$0.&000\,004\,85\,(12)  & $-$0.&000\,002\,51\,(6)  \\
\hline 
Total value $g$                       &    0.&652\,575\,5\,(20) &    1.&323\,705\,1\,(16) \\
\hline \hline
\end{tabular}
\end{center}
\end{table}

\begin{table}
\begin{center}
\begin{tabular}{lr@{}lr@{}l}
${}^{132}_{54}$Xe$^{51+}$ \\
\hline  
& \multicolumn{2}{c}{$(1s)^2 2p_{1/2}$}
& \multicolumn{2}{c}{$(1s)^2 2p_{3/2}$}
\\[6pt]
\hline 
Dirac value $\gdirac$                     &    0.&639\,416\,981     &    1.&312\,424\,274     \\
Interelectronic interaction $\dgint$  &    0.&001\,748\,8\,(37)  &    0.&001\,349\,5\,(29)  \\
One-loop QED $\dgqed$               & $-$0.&000\,689\,5\,(10)  & 0.&000\,840\,43\,(70) \\
Two-loop QED $\dgqed$               &    0.&000\,001\,2\,(1)  &    0.&000\,001\,2\,(1)  \\
Nuclear recoil $\dgnuc$               & $-$0.&000\,003\,23\,(6)  & $-$0.&000\,001\,72\,(3)  \\
\hline 
Total value $g$                       &    0.&640\,474\,2\,(38) &    1.&314\,613\,7\,(30) \\
\hline
\\
${}^{173}_{70}$Yb$^{67+}$ \\
\hline 
& \multicolumn{2}{c}{$(1s)^2 2p_{1/2}$}
& \multicolumn{2}{c}{$(1s)^2 2p_{3/2}$}
\\[6pt]
\hline 
Dirac value $\gdirac$                     &    0.&619\,047\,843     &    1.&297\,955\,774     \\
Interelectronic interaction $\dgint$  &    0.&002\,411\,8\,(69)  &    0.&001\,781\,6\,(57)  \\
One-loop QED $\dgqed$               & $-$0.&000\,598\,1\,(17)  & 0.&000\,895\,1\,(10) \\
Two-loop QED $\dgqed$               &    0.&000\,001\,2\,(1)  &    0.&000\,001\,2\,(1)  \\
Nuclear recoil $\dgnuc$               & $-$0.&000\,002\,36\,(3)  & $-$0.&000\,001\,32\,(2)  \\
\hline 
Total value $g$                       &    0.&620\,860\,0\,(71) &    1.&300\,631\,8\,(58) \\
\hline
\\
${}^{208}_{82}$Pb$^{79+}$ \\
\hline 
& \multicolumn{2}{c}{$(1s)^2 2p_{1/2}$}
& \multicolumn{2}{c}{$(1s)^2 2p_{3/2}$}
\\[6pt]
\hline  
Dirac value $\gdirac$                     &    0.&598\,676\,352     &    1.&284\,472\,641     \\
Interelectronic interaction $\dgint$  &    0.&003\,009\,(10)  &    0.&002\,126\,1\,(88)  \\
One-loop QED $\dgqed$               & $-$0.&000\,493\,0\,(24)  & 0.&000\,949\,5\,(12) \\
Two-loop QED $\dgqed$               &    0.&000\,001\,2\,(8)  &    0.&000\,001\,2\,(8)  \\
Nuclear recoil $\dgnuc$               & $-$0.&000\,001\,89\,(2)  & $-$0.&000\,001\,12\,(1)  \\
\hline 
Total value $g$                       &    0.&601\,189\,(10) &    1.&287\,544\,1\,(89) \\
\hline \hline
\end{tabular}
\end{center}
\end{table}

\begin{table}
\begin{center}
\begin{tabular}{lr@{}lr@{}l}
${}^{238}_{92}$U$^{89+}$ \\
\hline 
& \multicolumn{2}{c}{$(1s)^2 2p_{1/2}$}
& \multicolumn{2}{c}{$(1s)^2 2p_{3/2}$}
\\[6pt]
\hline 
Dirac value $\gdirac$                     &    0.&577\,417\,772     &    1.&271\,441\,831     \\
Interelectronic interaction $\dgint$  &    0.&003\,610\,(14)  &    0.&002\,433\,(12)  \\
One-loop QED $\dgqed$               & $-$0.&000\,380\,7\,(30)  & 0.&000\,992\,6\,(13) \\
Two-loop QED $\dgqed$               &    0.&000\,001\,2\,(8)  &    0.&000\,001\,2\,(8)  \\
Nuclear recoil $\dgnuc$               & $-$0.&000\,001\,56\,(2)  & $-$0.&000\,000\,99\,(1)  \\
\hline 
Total value $g$                       &    0.&580\,647\,(14) &    1.&274\,868\,(12) \\
\hline
\end{tabular}
\end{center}
\end{table}

%
\section{CONCLUSION}
We have presented theoretical values of the $g$ factor for the excited states of lithium-like ions over a wide range of nuclear charges, $Z = 10$–$92$, with an estimated uncertainty of the order of $10^{-6}$. The interelectronic interaction contributions have been considered up to the second order. The first-order correction has been obtained within the rigorous framework of bound-state QED, while the second-order term is treated using the Breit approximation. The one-loop QED corrections, including both self-energy and vacuum polarization contributions, are calculated to all orders in $\alpha Z$. The two-loop contributions are included via an $\alpha Z$ expansion. Additionally, we have incorporated nuclear recoil effects and calculated the leading nonlinear Zeeman contributions. This work represents the first complete relativistic treatment of the excited-state $g$ factor in lithium-like ions, offering a reliable theoretical foundation for forthcoming high-precision experimental studies.
%

%
\section*{Acknowledgments}
The research was supported by the Foundation for the Advancement of Theoretical Physics and Mathematics “BASIS”, by the Ministry of Science and Higher Education of the Russian Federation (Project No. FSER-2025–0012) and by the Russian Science Foundation (Project No. 22-12-00258).
%
%
%
\bibliographystyle{apsrev4-1}
\bibliography{blaze}

\begin{thebibliography}{69}%
\makeatletter
\providecommand \@ifxundefined [1]{%
 \@ifx{#1\undefined}
}%
\providecommand \@ifnum [1]{%
 \ifnum #1\expandafter \@firstoftwo
 \else \expandafter \@secondoftwo
 \fi
}%
\providecommand \@ifx [1]{%
 \ifx #1\expandafter \@firstoftwo
 \else \expandafter \@secondoftwo
 \fi
}%
\providecommand \natexlab [1]{#1}%
\providecommand \enquote  [1]{``#1''}%
\providecommand \bibnamefont  [1]{#1}%
\providecommand \bibfnamefont [1]{#1}%
\providecommand \citenamefont [1]{#1}%
\providecommand \href@noop [0]{\@secondoftwo}%
\providecommand \href [0]{\begingroup \@sanitize@url \@href}%
\providecommand \@href[1]{\@@startlink{#1}\@@href}%
\providecommand \@@href[1]{\endgroup#1\@@endlink}%
\providecommand \@sanitize@url [0]{\catcode `\\12\catcode `\$12\catcode `\&12\catcode `\#12\catcode `\^12\catcode `\_12\catcode `\%12\relax}%
\providecommand \@@startlink[1]{}%
\providecommand \@@endlink[0]{}%
\providecommand \url  [0]{\begingroup\@sanitize@url \@url }%
\providecommand \@url [1]{\endgroup\@href {#1}{\urlprefix }}%
\providecommand \urlprefix  [0]{URL }%
\providecommand \Eprint [0]{\href }%
\providecommand \doibase [0]{http://dx.doi.org/}%
\providecommand \selectlanguage [0]{\@gobble}%
\providecommand \bibinfo  [0]{\@secondoftwo}%
\providecommand \bibfield  [0]{\@secondoftwo}%
\providecommand \translation [1]{[#1]}%
\providecommand \BibitemOpen [0]{}%
\providecommand \bibitemStop [0]{}%
\providecommand \bibitemNoStop [0]{.\EOS\space}%
\providecommand \EOS [0]{\spacefactor3000\relax}%
\providecommand \BibitemShut  [1]{\csname bibitem#1\endcsname}%
\let\auto@bib@innerbib\@empty
\bibitem [{\citenamefont {Shabaev}\ \emph {et~al.}(2015)\citenamefont {Shabaev}, \citenamefont {Glazov}, \citenamefont {Plunien},\ and\ \citenamefont {Volotka}}]{Shabaev:2015:031205}%
  \BibitemOpen
  \bibfield  {author} {\bibinfo {author} {\bibfnamefont {V.~M.}\ \bibnamefont {Shabaev}}, \bibinfo {author} {\bibfnamefont {D.~A.}\ \bibnamefont {Glazov}}, \bibinfo {author} {\bibfnamefont {G.}~\bibnamefont {Plunien}}, \ and\ \bibinfo {author} {\bibfnamefont {A.~V.}\ \bibnamefont {Volotka}},\ }\href {\doibase 10.1063/1.4921299} {\bibfield  {journal} {\bibinfo  {journal} {Journal of Physical and Chemical Reference Data}\ }\textbf {\bibinfo {volume} {44}},\ \bibinfo {pages} {031205} (\bibinfo {year} {2015})}\BibitemShut {NoStop}%
\bibitem [{\citenamefont {Glazov}\ \emph {et~al.}(2023)\citenamefont {Glazov}, \citenamefont {Zinenko}, \citenamefont {Agababaev}, \citenamefont {Moshkin}, \citenamefont {Tryapitsyna}, \citenamefont {Volchkova},\ and\ \citenamefont {Volotka}}]{Glazov:2023:A}%
  \BibitemOpen
  \bibfield  {author} {\bibinfo {author} {\bibfnamefont {D.~A.}\ \bibnamefont {Glazov}}, \bibinfo {author} {\bibfnamefont {D.~V.}\ \bibnamefont {Zinenko}}, \bibinfo {author} {\bibfnamefont {V.~A.}\ \bibnamefont {Agababaev}}, \bibinfo {author} {\bibfnamefont {A.~D.}\ \bibnamefont {Moshkin}}, \bibinfo {author} {\bibfnamefont {E.~V.}\ \bibnamefont {Tryapitsyna}}, \bibinfo {author} {\bibfnamefont {A.~M.}\ \bibnamefont {Volchkova}}, \ and\ \bibinfo {author} {\bibfnamefont {A.~V.}\ \bibnamefont {Volotka}},\ }\href {\doibase 10.3390/atoms11090119} {\bibfield  {journal} {\bibinfo  {journal} {Atoms}\ }\textbf {\bibinfo {volume} {11}} (\bibinfo {year} {2023}),\ 10.3390/atoms11090119}\BibitemShut {NoStop}%
\bibitem [{\citenamefont {Sturm}\ \emph {et~al.}(2014)\citenamefont {Sturm}, \citenamefont {K{\"o}hler}, \citenamefont {Zatorski}, \citenamefont {Wagner}, \citenamefont {Harman}, \citenamefont {Werth}, \citenamefont {Quint}, \citenamefont {Keitel},\ and\ \citenamefont {Blaum}}]{sturm:2014:467}%
  \BibitemOpen
  \bibfield  {author} {\bibinfo {author} {\bibfnamefont {S.}~\bibnamefont {Sturm}}, \bibinfo {author} {\bibfnamefont {F.}~\bibnamefont {K{\"o}hler}}, \bibinfo {author} {\bibfnamefont {J.}~\bibnamefont {Zatorski}}, \bibinfo {author} {\bibfnamefont {A.}~\bibnamefont {Wagner}}, \bibinfo {author} {\bibfnamefont {Z.}~\bibnamefont {Harman}}, \bibinfo {author} {\bibfnamefont {G.}~\bibnamefont {Werth}}, \bibinfo {author} {\bibfnamefont {W.}~\bibnamefont {Quint}}, \bibinfo {author} {\bibfnamefont {C.~H.}\ \bibnamefont {Keitel}}, \ and\ \bibinfo {author} {\bibfnamefont {K.}~\bibnamefont {Blaum}},\ }\href {\doibase 10.1038/nature13026} {\bibfield  {journal} {\bibinfo  {journal} {Nature}\ }\textbf {\bibinfo {volume} {506}},\ \bibinfo {pages} {467} (\bibinfo {year} {2014})}\BibitemShut {NoStop}%
\bibitem [{\citenamefont {Mohr}\ \emph {et~al.}(2025)\citenamefont {Mohr}, \citenamefont {Newell}, \citenamefont {Taylor},\ and\ \citenamefont {Tiesinga}}]{CODATA:2025:025002}%
  \BibitemOpen
  \bibfield  {author} {\bibinfo {author} {\bibfnamefont {P.~J.}\ \bibnamefont {Mohr}}, \bibinfo {author} {\bibfnamefont {D.~B.}\ \bibnamefont {Newell}}, \bibinfo {author} {\bibfnamefont {B.~N.}\ \bibnamefont {Taylor}}, \ and\ \bibinfo {author} {\bibfnamefont {E.}~\bibnamefont {Tiesinga}},\ }\href {\doibase 10.1103/RevModPhys.97.025002} {\bibfield  {journal} {\bibinfo  {journal} {Rev. Mod. Phys.}\ }\textbf {\bibinfo {volume} {97}},\ \bibinfo {pages} {025002} (\bibinfo {year} {2025})}\BibitemShut {NoStop}%
\bibitem [{\citenamefont {Shabaev}\ \emph {et~al.}(2006)\citenamefont {Shabaev}, \citenamefont {Glazov}, \citenamefont {Oreshkina}, \citenamefont {Volotka}, \citenamefont {Plunien}, \citenamefont {Kluge},\ and\ \citenamefont {Quint}}]{Shabaev:2006:253002}%
  \BibitemOpen
  \bibfield  {author} {\bibinfo {author} {\bibfnamefont {V.~M.}\ \bibnamefont {Shabaev}}, \bibinfo {author} {\bibfnamefont {D.~A.}\ \bibnamefont {Glazov}}, \bibinfo {author} {\bibfnamefont {N.~S.}\ \bibnamefont {Oreshkina}}, \bibinfo {author} {\bibfnamefont {A.~V.}\ \bibnamefont {Volotka}}, \bibinfo {author} {\bibfnamefont {G.}~\bibnamefont {Plunien}}, \bibinfo {author} {\bibfnamefont {H.-J.}\ \bibnamefont {Kluge}}, \ and\ \bibinfo {author} {\bibfnamefont {W.}~\bibnamefont {Quint}},\ }\href {\doibase 10.1103/PhysRevLett.96.253002} {\bibfield  {journal} {\bibinfo  {journal} {Physical Review Letters}\ }\textbf {\bibinfo {volume} {96}},\ \bibinfo {pages} {253002} (\bibinfo {year} {2006})}\BibitemShut {NoStop}%
\bibitem [{\citenamefont {Yerokhin}\ \emph {et~al.}(2016)\citenamefont {Yerokhin}, \citenamefont {Berseneva}, \citenamefont {Harman}, \citenamefont {Tupitsyn},\ and\ \citenamefont {Keitel}}]{Yerokhin:2016:100801}%
  \BibitemOpen
  \bibfield  {author} {\bibinfo {author} {\bibfnamefont {V.~A.}\ \bibnamefont {Yerokhin}}, \bibinfo {author} {\bibfnamefont {E.}~\bibnamefont {Berseneva}}, \bibinfo {author} {\bibfnamefont {Z.}~\bibnamefont {Harman}}, \bibinfo {author} {\bibfnamefont {I.~I.}\ \bibnamefont {Tupitsyn}}, \ and\ \bibinfo {author} {\bibfnamefont {C.~H.}\ \bibnamefont {Keitel}},\ }\href {\doibase 10.1103/PhysRevLett.116.100801} {\bibfield  {journal} {\bibinfo  {journal} {Physical Review Letters}\ }\textbf {\bibinfo {volume} {116}},\ \bibinfo {pages} {100801} (\bibinfo {year} {2016})}\BibitemShut {NoStop}%
\bibitem [{\citenamefont {Shabaev}\ \emph {et~al.}(2017)\citenamefont {Shabaev}, \citenamefont {Glazov}, \citenamefont {Malyshev},\ and\ \citenamefont {Tupitsyn}}]{Shabaev:2017:263001}%
  \BibitemOpen
  \bibfield  {author} {\bibinfo {author} {\bibfnamefont {V.~M.}\ \bibnamefont {Shabaev}}, \bibinfo {author} {\bibfnamefont {D.~A.}\ \bibnamefont {Glazov}}, \bibinfo {author} {\bibfnamefont {A.~V.}\ \bibnamefont {Malyshev}}, \ and\ \bibinfo {author} {\bibfnamefont {I.~I.}\ \bibnamefont {Tupitsyn}},\ }\href {\doibase 10.1103/PhysRevLett.119.263001} {\bibfield  {journal} {\bibinfo  {journal} {Physical Review Letters}\ }\textbf {\bibinfo {volume} {119}},\ \bibinfo {pages} {263001} (\bibinfo {year} {2017})}\BibitemShut {NoStop}%
\bibitem [{\citenamefont {Malyshev}\ \emph {et~al.}(2017)\citenamefont {Malyshev}, \citenamefont {Shabaev}, \citenamefont {Glazov},\ and\ \citenamefont {Tupitsyn}}]{Malyshev:2017:765}%
  \BibitemOpen
  \bibfield  {author} {\bibinfo {author} {\bibfnamefont {A.~V.}\ \bibnamefont {Malyshev}}, \bibinfo {author} {\bibfnamefont {V.~M.}\ \bibnamefont {Shabaev}}, \bibinfo {author} {\bibfnamefont {D.~A.}\ \bibnamefont {Glazov}}, \ and\ \bibinfo {author} {\bibfnamefont {I.~I.}\ \bibnamefont {Tupitsyn}},\ }\href {\doibase 10.1134/S0021364017240018} {\bibfield  {journal} {\bibinfo  {journal} {JETP Letters}\ }\textbf {\bibinfo {volume} {106}},\ \bibinfo {pages} {765} (\bibinfo {year} {2017})}\BibitemShut {NoStop}%
\bibitem [{\citenamefont {Shabaev}\ \emph {et~al.}(2018)\citenamefont {Shabaev}, \citenamefont {Glazov}, \citenamefont {Malyshev},\ and\ \citenamefont {Tupitsyn}}]{Shabaev:2018:032512}%
  \BibitemOpen
  \bibfield  {author} {\bibinfo {author} {\bibfnamefont {V.~M.}\ \bibnamefont {Shabaev}}, \bibinfo {author} {\bibfnamefont {D.~A.}\ \bibnamefont {Glazov}}, \bibinfo {author} {\bibfnamefont {A.~V.}\ \bibnamefont {Malyshev}}, \ and\ \bibinfo {author} {\bibfnamefont {I.~I.}\ \bibnamefont {Tupitsyn}},\ }\href {\doibase 10.1103/PhysRevA.98.032512} {\bibfield  {journal} {\bibinfo  {journal} {Physical Review A}\ }\textbf {\bibinfo {volume} {98}},\ \bibinfo {pages} {032512} (\bibinfo {year} {2018})}\BibitemShut {NoStop}%
\bibitem [{\citenamefont {Quint}\ \emph {et~al.}(2008)\citenamefont {Quint}, \citenamefont {Moskovkhin}, \citenamefont {Shabaev},\ and\ \citenamefont {Vogel}}]{Quint:2008:032517}%
  \BibitemOpen
  \bibfield  {author} {\bibinfo {author} {\bibfnamefont {W.}~\bibnamefont {Quint}}, \bibinfo {author} {\bibfnamefont {D.~L.}\ \bibnamefont {Moskovkhin}}, \bibinfo {author} {\bibfnamefont {V.~M.}\ \bibnamefont {Shabaev}}, \ and\ \bibinfo {author} {\bibfnamefont {M.}~\bibnamefont {Vogel}},\ }\href {\doibase 10.1103/PhysRevA.78.032517} {\bibfield  {journal} {\bibinfo  {journal} {Physical Review A}\ }\textbf {\bibinfo {volume} {78}},\ \bibinfo {pages} {032517} (\bibinfo {year} {2008})}\BibitemShut {NoStop}%
\bibitem [{\citenamefont {Yerokhin}\ \emph {et~al.}(2011)\citenamefont {Yerokhin}, \citenamefont {Pachucki}, \citenamefont {Harman},\ and\ \citenamefont {Keitel}}]{Yerokhin:2011:043004}%
  \BibitemOpen
  \bibfield  {author} {\bibinfo {author} {\bibfnamefont {V.~A.}\ \bibnamefont {Yerokhin}}, \bibinfo {author} {\bibfnamefont {K.}~\bibnamefont {Pachucki}}, \bibinfo {author} {\bibfnamefont {Z.}~\bibnamefont {Harman}}, \ and\ \bibinfo {author} {\bibfnamefont {C.~H.}\ \bibnamefont {Keitel}},\ }\href {\doibase 10.1103/PhysRevLett.107.043004} {\bibfield  {journal} {\bibinfo  {journal} {Physical Review Letters}\ }\textbf {\bibinfo {volume} {107}},\ \bibinfo {pages} {043004} (\bibinfo {year} {2011})}\BibitemShut {NoStop}%
\bibitem [{\citenamefont {Sailer}\ \emph {et~al.}(2022)\citenamefont {Sailer}, \citenamefont {Debierre}, \citenamefont {Harman}, \citenamefont {Hei{\ss}e}, \citenamefont {K{\"o}nig}, \citenamefont {Morgner}, \citenamefont {Tu}, \citenamefont {Volotka}, \citenamefont {Keitel}, \citenamefont {Blaum},\ and\ \citenamefont {Sturm}}]{Sailer:2022:479}%
  \BibitemOpen
  \bibfield  {author} {\bibinfo {author} {\bibfnamefont {T.}~\bibnamefont {Sailer}}, \bibinfo {author} {\bibfnamefont {V.}~\bibnamefont {Debierre}}, \bibinfo {author} {\bibfnamefont {Z.}~\bibnamefont {Harman}}, \bibinfo {author} {\bibfnamefont {F.}~\bibnamefont {Hei{\ss}e}}, \bibinfo {author} {\bibfnamefont {C.}~\bibnamefont {K{\"o}nig}}, \bibinfo {author} {\bibfnamefont {J.}~\bibnamefont {Morgner}}, \bibinfo {author} {\bibfnamefont {B.}~\bibnamefont {Tu}}, \bibinfo {author} {\bibfnamefont {A.~V.}\ \bibnamefont {Volotka}}, \bibinfo {author} {\bibfnamefont {C.~H.}\ \bibnamefont {Keitel}}, \bibinfo {author} {\bibfnamefont {K.}~\bibnamefont {Blaum}}, \ and\ \bibinfo {author} {\bibfnamefont {S.}~\bibnamefont {Sturm}},\ }\href {\doibase 10.1038/s41586-022-04807-w} {\bibfield  {journal} {\bibinfo  {journal} {Nature}\ }\textbf {\bibinfo {volume} {606}},\ \bibinfo {pages} {479} (\bibinfo {year} {2022})}\BibitemShut {NoStop}%
\bibitem [{\citenamefont {Shabaev}\ \emph {et~al.}(2022)\citenamefont {Shabaev}, \citenamefont {Glazov}, \citenamefont {Ryzhkov}, \citenamefont {Brandau}, \citenamefont {Plunien}, \citenamefont {Quint}, \citenamefont {Volchkova},\ and\ \citenamefont {Zinenko}}]{shabaev:2022:043001}%
  \BibitemOpen
  \bibfield  {author} {\bibinfo {author} {\bibfnamefont {V.~M.}\ \bibnamefont {Shabaev}}, \bibinfo {author} {\bibfnamefont {D.~A.}\ \bibnamefont {Glazov}}, \bibinfo {author} {\bibfnamefont {A.~M.}\ \bibnamefont {Ryzhkov}}, \bibinfo {author} {\bibfnamefont {C.}~\bibnamefont {Brandau}}, \bibinfo {author} {\bibfnamefont {G.}~\bibnamefont {Plunien}}, \bibinfo {author} {\bibfnamefont {W.}~\bibnamefont {Quint}}, \bibinfo {author} {\bibfnamefont {A.~M.}\ \bibnamefont {Volchkova}}, \ and\ \bibinfo {author} {\bibfnamefont {D.~V.}\ \bibnamefont {Zinenko}},\ }\href {\doibase 10.1103/PhysRevLett.128.043001} {\bibfield  {journal} {\bibinfo  {journal} {Physical Review Letters}\ }\textbf {\bibinfo {volume} {128}},\ \bibinfo {pages} {043001} (\bibinfo {year} {2022})}\BibitemShut {NoStop}%
\bibitem [{\citenamefont {Peik}\ and\ \citenamefont {Tamm}(2003)}]{Peik:2003:181}%
  \BibitemOpen
  \bibfield  {author} {\bibinfo {author} {\bibfnamefont {E.}~\bibnamefont {Peik}}\ and\ \bibinfo {author} {\bibfnamefont {C.}~\bibnamefont {Tamm}},\ }\href {\doibase 10.1209/epl/i2003-00210-x} {\bibfield  {journal} {\bibinfo  {journal} {Europhysics Letters}\ }\textbf {\bibinfo {volume} {61}},\ \bibinfo {pages} {181} (\bibinfo {year} {2003})}\BibitemShut {NoStop}%
\bibitem [{\citenamefont {Campbell}\ \emph {et~al.}(2012)\citenamefont {Campbell}, \citenamefont {Radnaev}, \citenamefont {Kuzmich}, \citenamefont {Dzuba}, \citenamefont {Flambaum},\ and\ \citenamefont {Derevianko}}]{Campbell:2012:120802}%
  \BibitemOpen
  \bibfield  {author} {\bibinfo {author} {\bibfnamefont {C.~J.}\ \bibnamefont {Campbell}}, \bibinfo {author} {\bibfnamefont {A.~G.}\ \bibnamefont {Radnaev}}, \bibinfo {author} {\bibfnamefont {A.}~\bibnamefont {Kuzmich}}, \bibinfo {author} {\bibfnamefont {V.~A.}\ \bibnamefont {Dzuba}}, \bibinfo {author} {\bibfnamefont {V.~V.}\ \bibnamefont {Flambaum}}, \ and\ \bibinfo {author} {\bibfnamefont {A.}~\bibnamefont {Derevianko}},\ }\href {\doibase 10.1103/PhysRevLett.108.120802} {\bibfield  {journal} {\bibinfo  {journal} {Physical Review Letters}\ }\textbf {\bibinfo {volume} {108}},\ \bibinfo {pages} {120802} (\bibinfo {year} {2012})}\BibitemShut {NoStop}%
\bibitem [{\citenamefont {Tkalya}(2011)}]{Tkalya:2011:162501}%
  \BibitemOpen
  \bibfield  {author} {\bibinfo {author} {\bibfnamefont {E.~V.}\ \bibnamefont {Tkalya}},\ }\href {\doibase 10.1103/PhysRevLett.106.162501} {\bibfield  {journal} {\bibinfo  {journal} {Physical Review Letters}\ }\textbf {\bibinfo {volume} {106}},\ \bibinfo {pages} {162501} (\bibinfo {year} {2011})}\BibitemShut {NoStop}%
\bibitem [{\citenamefont {Schiffmann}\ \emph {et~al.}(2021)\citenamefont {Schiffmann}, \citenamefont {Brage}, \citenamefont {Judge}, \citenamefont {Paraschiv},\ and\ \citenamefont {Wang}}]{Schiffmann:2021:186}%
  \BibitemOpen
  \bibfield  {author} {\bibinfo {author} {\bibfnamefont {S.}~\bibnamefont {Schiffmann}}, \bibinfo {author} {\bibfnamefont {T.}~\bibnamefont {Brage}}, \bibinfo {author} {\bibfnamefont {P.~G.}\ \bibnamefont {Judge}}, \bibinfo {author} {\bibfnamefont {A.~R.}\ \bibnamefont {Paraschiv}}, \ and\ \bibinfo {author} {\bibfnamefont {K.}~\bibnamefont {Wang}},\ }\href {\doibase 10.3847/1538-4357/ac2cca} {\bibfield  {journal} {\bibinfo  {journal} {The Astrophysical Journal}\ }\textbf {\bibinfo {volume} {923}},\ \bibinfo {pages} {186} (\bibinfo {year} {2021})}\BibitemShut {NoStop}%
\bibitem [{\citenamefont {H{\"a}ffner}\ \emph {et~al.}(2000)\citenamefont {H{\"a}ffner}, \citenamefont {Beier}, \citenamefont {Hermanspahn}, \citenamefont {Kluge}, \citenamefont {Quint}, \citenamefont {Stahl}, \citenamefont {Verd{\'u}},\ and\ \citenamefont {Werth}}]{Haffner:2000:5308}%
  \BibitemOpen
  \bibfield  {author} {\bibinfo {author} {\bibfnamefont {H.}~\bibnamefont {H{\"a}ffner}}, \bibinfo {author} {\bibfnamefont {T.}~\bibnamefont {Beier}}, \bibinfo {author} {\bibfnamefont {N.}~\bibnamefont {Hermanspahn}}, \bibinfo {author} {\bibfnamefont {H.-J.}\ \bibnamefont {Kluge}}, \bibinfo {author} {\bibfnamefont {W.}~\bibnamefont {Quint}}, \bibinfo {author} {\bibfnamefont {S.}~\bibnamefont {Stahl}}, \bibinfo {author} {\bibfnamefont {J.}~\bibnamefont {Verd{\'u}}}, \ and\ \bibinfo {author} {\bibfnamefont {G.}~\bibnamefont {Werth}},\ }\href {\doibase 10.1103/PhysRevLett.85.5308} {\bibfield  {journal} {\bibinfo  {journal} {Physical Review Letters}\ }\textbf {\bibinfo {volume} {85}},\ \bibinfo {pages} {5308} (\bibinfo {year} {2000})}\BibitemShut {NoStop}%
\bibitem [{\citenamefont {Verd{\'u}}\ \emph {et~al.}(2004)\citenamefont {Verd{\'u}}, \citenamefont {Djeki{\'c}}, \citenamefont {Stahl}, \citenamefont {Valenzuela}, \citenamefont {Vogel}, \citenamefont {Werth}, \citenamefont {Beier}, \citenamefont {Kluge},\ and\ \citenamefont {Quint}}]{Verdu:2004:093002}%
  \BibitemOpen
  \bibfield  {author} {\bibinfo {author} {\bibfnamefont {J.}~\bibnamefont {Verd{\'u}}}, \bibinfo {author} {\bibfnamefont {S.}~\bibnamefont {Djeki{\'c}}}, \bibinfo {author} {\bibfnamefont {S.}~\bibnamefont {Stahl}}, \bibinfo {author} {\bibfnamefont {T.}~\bibnamefont {Valenzuela}}, \bibinfo {author} {\bibfnamefont {M.}~\bibnamefont {Vogel}}, \bibinfo {author} {\bibfnamefont {G.}~\bibnamefont {Werth}}, \bibinfo {author} {\bibfnamefont {T.}~\bibnamefont {Beier}}, \bibinfo {author} {\bibfnamefont {H.-J.}\ \bibnamefont {Kluge}}, \ and\ \bibinfo {author} {\bibfnamefont {W.}~\bibnamefont {Quint}},\ }\href {\doibase 10.1103/PhysRevLett.92.093002} {\bibfield  {journal} {\bibinfo  {journal} {Physical Review Letters}\ }\textbf {\bibinfo {volume} {92}},\ \bibinfo {pages} {093002} (\bibinfo {year} {2004})}\BibitemShut {NoStop}%
\bibitem [{\citenamefont {Sturm}\ \emph {et~al.}(2011)\citenamefont {Sturm}, \citenamefont {Wagner}, \citenamefont {Schabinger}, \citenamefont {Zatorski}, \citenamefont {Harman}, \citenamefont {Quint}, \citenamefont {Werth}, \citenamefont {Keitel},\ and\ \citenamefont {Blaum}}]{Sturm:2011:023002}%
  \BibitemOpen
  \bibfield  {author} {\bibinfo {author} {\bibfnamefont {S.}~\bibnamefont {Sturm}}, \bibinfo {author} {\bibfnamefont {A.}~\bibnamefont {Wagner}}, \bibinfo {author} {\bibfnamefont {B.}~\bibnamefont {Schabinger}}, \bibinfo {author} {\bibfnamefont {J.}~\bibnamefont {Zatorski}}, \bibinfo {author} {\bibfnamefont {Z.}~\bibnamefont {Harman}}, \bibinfo {author} {\bibfnamefont {W.}~\bibnamefont {Quint}}, \bibinfo {author} {\bibfnamefont {G.}~\bibnamefont {Werth}}, \bibinfo {author} {\bibfnamefont {C.~H.}\ \bibnamefont {Keitel}}, \ and\ \bibinfo {author} {\bibfnamefont {K.}~\bibnamefont {Blaum}},\ }\href {\doibase 10.1103/PhysRevLett.107.023002} {\bibfield  {journal} {\bibinfo  {journal} {Physical Review Letters}\ }\textbf {\bibinfo {volume} {107}},\ \bibinfo {pages} {023002} (\bibinfo {year} {2011})}\BibitemShut {NoStop}%
\bibitem [{\citenamefont {Sturm}\ \emph {et~al.}(2013)\citenamefont {Sturm}, \citenamefont {Wagner}, \citenamefont {Kretzschmar}, \citenamefont {Quint}, \citenamefont {Werth},\ and\ \citenamefont {Blaum}}]{Sturm:2013:030501}%
  \BibitemOpen
  \bibfield  {author} {\bibinfo {author} {\bibfnamefont {S.}~\bibnamefont {Sturm}}, \bibinfo {author} {\bibfnamefont {A.}~\bibnamefont {Wagner}}, \bibinfo {author} {\bibfnamefont {M.}~\bibnamefont {Kretzschmar}}, \bibinfo {author} {\bibfnamefont {W.}~\bibnamefont {Quint}}, \bibinfo {author} {\bibfnamefont {G.}~\bibnamefont {Werth}}, \ and\ \bibinfo {author} {\bibfnamefont {K.}~\bibnamefont {Blaum}},\ }\href {\doibase 10.1103/PhysRevA.87.030501} {\bibfield  {journal} {\bibinfo  {journal} {Physical Review A}\ }\textbf {\bibinfo {volume} {87}},\ \bibinfo {pages} {030501} (\bibinfo {year} {2013})}\BibitemShut {NoStop}%
\bibitem [{\citenamefont {Hei\ss{}e}\ \emph {et~al.}(2023)\citenamefont {Hei\ss{}e}, \citenamefont {Door}, \citenamefont {Sailer}, \citenamefont {Filianin}, \citenamefont {Herkenhoff}, \citenamefont {K\"onig}, \citenamefont {Kromer}, \citenamefont {Lange}, \citenamefont {Morgner}, \citenamefont {Rischka}, \citenamefont {Schweiger}, \citenamefont {Tu}, \citenamefont {Novikov}, \citenamefont {Eliseev}, \citenamefont {Sturm},\ and\ \citenamefont {Blaum}}]{Heisse:2023:PRL}%
  \BibitemOpen
  \bibfield  {author} {\bibinfo {author} {\bibfnamefont {F.}~\bibnamefont {Hei\ss{}e}}, \bibinfo {author} {\bibfnamefont {M.}~\bibnamefont {Door}}, \bibinfo {author} {\bibfnamefont {T.}~\bibnamefont {Sailer}}, \bibinfo {author} {\bibfnamefont {P.}~\bibnamefont {Filianin}}, \bibinfo {author} {\bibfnamefont {J.}~\bibnamefont {Herkenhoff}}, \bibinfo {author} {\bibfnamefont {C.~M.}\ \bibnamefont {K\"onig}}, \bibinfo {author} {\bibfnamefont {K.}~\bibnamefont {Kromer}}, \bibinfo {author} {\bibfnamefont {D.}~\bibnamefont {Lange}}, \bibinfo {author} {\bibfnamefont {J.}~\bibnamefont {Morgner}}, \bibinfo {author} {\bibfnamefont {A.}~\bibnamefont {Rischka}}, \bibinfo {author} {\bibfnamefont {C.}~\bibnamefont {Schweiger}}, \bibinfo {author} {\bibfnamefont {B.}~\bibnamefont {Tu}}, \bibinfo {author} {\bibfnamefont {Y.~N.}\ \bibnamefont {Novikov}}, \bibinfo {author} {\bibfnamefont {S.}~\bibnamefont {Eliseev}}, \bibinfo {author} {\bibfnamefont {S.}~\bibnamefont {Sturm}}, \ and\ \bibinfo {author} {\bibfnamefont
  {K.}~\bibnamefont {Blaum}},\ }\href {\doibase 10.1103/PhysRevLett.131.253002} {\bibfield  {journal} {\bibinfo  {journal} {Phys. Rev. Lett.}\ }\textbf {\bibinfo {volume} {131}},\ \bibinfo {pages} {253002} (\bibinfo {year} {2023})}\BibitemShut {NoStop}%
\bibitem [{\citenamefont {Morgner}\ \emph {et~al.}(2023)\citenamefont {Morgner}, \citenamefont {Tu}, \citenamefont {K{\"o}nig}, \citenamefont {Sailer}, \citenamefont {Hei{\ss}e}, \citenamefont {Bekker}, \citenamefont {Sikora}, \citenamefont {Lyu}, \citenamefont {Yerokhin}, \citenamefont {Harman}, \citenamefont {{Crespo L{\'o}pez-Urrutia}}, \citenamefont {Keitel}, \citenamefont {Sturm},\ and\ \citenamefont {Blaum}}]{Morgner:2023:53}%
  \BibitemOpen
  \bibfield  {author} {\bibinfo {author} {\bibfnamefont {J.}~\bibnamefont {Morgner}}, \bibinfo {author} {\bibfnamefont {B.}~\bibnamefont {Tu}}, \bibinfo {author} {\bibfnamefont {C.~M.}\ \bibnamefont {K{\"o}nig}}, \bibinfo {author} {\bibfnamefont {T.}~\bibnamefont {Sailer}}, \bibinfo {author} {\bibfnamefont {F.}~\bibnamefont {Hei{\ss}e}}, \bibinfo {author} {\bibfnamefont {H.}~\bibnamefont {Bekker}}, \bibinfo {author} {\bibfnamefont {B.}~\bibnamefont {Sikora}}, \bibinfo {author} {\bibfnamefont {C.}~\bibnamefont {Lyu}}, \bibinfo {author} {\bibfnamefont {V.~A.}\ \bibnamefont {Yerokhin}}, \bibinfo {author} {\bibfnamefont {Z.}~\bibnamefont {Harman}}, \bibinfo {author} {\bibfnamefont {J.~R.}\ \bibnamefont {{Crespo L{\'o}pez-Urrutia}}}, \bibinfo {author} {\bibfnamefont {C.~H.}\ \bibnamefont {Keitel}}, \bibinfo {author} {\bibfnamefont {S.}~\bibnamefont {Sturm}}, \ and\ \bibinfo {author} {\bibfnamefont {K.}~\bibnamefont {Blaum}},\ }\href {\doibase 10.1038/s41586-023-06453-2} {\bibfield  {journal} {\bibinfo  {journal}
  {Nature}\ }\textbf {\bibinfo {volume} {622}},\ \bibinfo {pages} {53} (\bibinfo {year} {2023})}\BibitemShut {NoStop}%
\bibitem [{\citenamefont {Shabaev}\ \emph {et~al.}(2002)\citenamefont {Shabaev}, \citenamefont {Glazov}, \citenamefont {Shabaeva}, \citenamefont {Yerokhin}, \citenamefont {Plunien},\ and\ \citenamefont {Soff}}]{Shabaev:2002:062104}%
  \BibitemOpen
  \bibfield  {author} {\bibinfo {author} {\bibfnamefont {V.~M.}\ \bibnamefont {Shabaev}}, \bibinfo {author} {\bibfnamefont {D.~A.}\ \bibnamefont {Glazov}}, \bibinfo {author} {\bibfnamefont {M.~B.}\ \bibnamefont {Shabaeva}}, \bibinfo {author} {\bibfnamefont {V.~A.}\ \bibnamefont {Yerokhin}}, \bibinfo {author} {\bibfnamefont {G.}~\bibnamefont {Plunien}}, \ and\ \bibinfo {author} {\bibfnamefont {G.}~\bibnamefont {Soff}},\ }\href {\doibase 10.1103/PhysRevA.65.062104} {\bibfield  {journal} {\bibinfo  {journal} {Physical Review A}\ }\textbf {\bibinfo {volume} {65}},\ \bibinfo {pages} {062104} (\bibinfo {year} {2002})}\BibitemShut {NoStop}%
\bibitem [{\citenamefont {Volotka}\ and\ \citenamefont {Plunien}(2014)}]{Volotka:2014:023002}%
  \BibitemOpen
  \bibfield  {author} {\bibinfo {author} {\bibfnamefont {A.~V.}\ \bibnamefont {Volotka}}\ and\ \bibinfo {author} {\bibfnamefont {G.}~\bibnamefont {Plunien}},\ }\href {\doibase 10.1103/PhysRevLett.113.023002} {\bibfield  {journal} {\bibinfo  {journal} {Physical Review Letters}\ }\textbf {\bibinfo {volume} {113}},\ \bibinfo {pages} {023002} (\bibinfo {year} {2014})}\BibitemShut {NoStop}%
\bibitem [{\citenamefont {Wagner}\ \emph {et~al.}(2013)\citenamefont {Wagner}, \citenamefont {Sturm}, \citenamefont {K{\"o}hler}, \citenamefont {Glazov}, \citenamefont {Volotka}, \citenamefont {Plunien}, \citenamefont {Quint}, \citenamefont {Werth}, \citenamefont {Shabaev},\ and\ \citenamefont {Blaum}}]{Wagner:2013:033003}%
  \BibitemOpen
  \bibfield  {author} {\bibinfo {author} {\bibfnamefont {A.}~\bibnamefont {Wagner}}, \bibinfo {author} {\bibfnamefont {S.}~\bibnamefont {Sturm}}, \bibinfo {author} {\bibfnamefont {F.}~\bibnamefont {K{\"o}hler}}, \bibinfo {author} {\bibfnamefont {D.~A.}\ \bibnamefont {Glazov}}, \bibinfo {author} {\bibfnamefont {A.~V.}\ \bibnamefont {Volotka}}, \bibinfo {author} {\bibfnamefont {G.}~\bibnamefont {Plunien}}, \bibinfo {author} {\bibfnamefont {W.}~\bibnamefont {Quint}}, \bibinfo {author} {\bibfnamefont {G.}~\bibnamefont {Werth}}, \bibinfo {author} {\bibfnamefont {V.~M.}\ \bibnamefont {Shabaev}}, \ and\ \bibinfo {author} {\bibfnamefont {K.}~\bibnamefont {Blaum}},\ }\href {\doibase 10.1103/PhysRevLett.110.033003} {\bibfield  {journal} {\bibinfo  {journal} {Physical Review Letters}\ }\textbf {\bibinfo {volume} {110}},\ \bibinfo {pages} {033003} (\bibinfo {year} {2013})}\BibitemShut {NoStop}%
\bibitem [{\citenamefont {Glazov}\ \emph {et~al.}(2019)\citenamefont {Glazov}, \citenamefont {{K{\"o}hler-Langes}}, \citenamefont {Volotka}, \citenamefont {Blaum}, \citenamefont {Hei{\ss}e}, \citenamefont {Plunien}, \citenamefont {Quint}, \citenamefont {Rau}, \citenamefont {Shabaev}, \citenamefont {Sturm},\ and\ \citenamefont {Werth}}]{Glazov:2019:173001}%
  \BibitemOpen
  \bibfield  {author} {\bibinfo {author} {\bibfnamefont {D.~A.}\ \bibnamefont {Glazov}}, \bibinfo {author} {\bibfnamefont {F.}~\bibnamefont {{K{\"o}hler-Langes}}}, \bibinfo {author} {\bibfnamefont {A.~V.}\ \bibnamefont {Volotka}}, \bibinfo {author} {\bibfnamefont {K.}~\bibnamefont {Blaum}}, \bibinfo {author} {\bibfnamefont {F.}~\bibnamefont {Hei{\ss}e}}, \bibinfo {author} {\bibfnamefont {G.}~\bibnamefont {Plunien}}, \bibinfo {author} {\bibfnamefont {W.}~\bibnamefont {Quint}}, \bibinfo {author} {\bibfnamefont {S.}~\bibnamefont {Rau}}, \bibinfo {author} {\bibfnamefont {V.~M.}\ \bibnamefont {Shabaev}}, \bibinfo {author} {\bibfnamefont {S.}~\bibnamefont {Sturm}}, \ and\ \bibinfo {author} {\bibfnamefont {G.}~\bibnamefont {Werth}},\ }\href {\doibase 10.1103/PhysRevLett.123.173001} {\bibfield  {journal} {\bibinfo  {journal} {Physical Review Letters}\ }\textbf {\bibinfo {volume} {123}},\ \bibinfo {pages} {173001} (\bibinfo {year} {2019})}\BibitemShut {NoStop}%
\bibitem [{\citenamefont {K{\"o}hler}\ \emph {et~al.}(2016)\citenamefont {K{\"o}hler}, \citenamefont {Blaum}, \citenamefont {Block}, \citenamefont {Chenmarev}, \citenamefont {Eliseev}, \citenamefont {Glazov}, \citenamefont {Goncharov}, \citenamefont {Hou}, \citenamefont {Kracke}, \citenamefont {Nesterenko}, \citenamefont {Novikov}, \citenamefont {Quint}, \citenamefont {Minaya~Ramirez}, \citenamefont {Shabaev}, \citenamefont {Sturm}, \citenamefont {Volotka},\ and\ \citenamefont {Werth}}]{Kohler:2016:10246}%
  \BibitemOpen
  \bibfield  {author} {\bibinfo {author} {\bibfnamefont {F.}~\bibnamefont {K{\"o}hler}}, \bibinfo {author} {\bibfnamefont {K.}~\bibnamefont {Blaum}}, \bibinfo {author} {\bibfnamefont {M.}~\bibnamefont {Block}}, \bibinfo {author} {\bibfnamefont {S.}~\bibnamefont {Chenmarev}}, \bibinfo {author} {\bibfnamefont {S.}~\bibnamefont {Eliseev}}, \bibinfo {author} {\bibfnamefont {D.~A.}\ \bibnamefont {Glazov}}, \bibinfo {author} {\bibfnamefont {M.}~\bibnamefont {Goncharov}}, \bibinfo {author} {\bibfnamefont {J.}~\bibnamefont {Hou}}, \bibinfo {author} {\bibfnamefont {A.}~\bibnamefont {Kracke}}, \bibinfo {author} {\bibfnamefont {D.~A.}\ \bibnamefont {Nesterenko}}, \bibinfo {author} {\bibfnamefont {Y.~N.}\ \bibnamefont {Novikov}}, \bibinfo {author} {\bibfnamefont {W.}~\bibnamefont {Quint}}, \bibinfo {author} {\bibfnamefont {E.}~\bibnamefont {Minaya~Ramirez}}, \bibinfo {author} {\bibfnamefont {V.~M.}\ \bibnamefont {Shabaev}}, \bibinfo {author} {\bibfnamefont {S.}~\bibnamefont {Sturm}}, \bibinfo {author} {\bibfnamefont
  {A.~V.}\ \bibnamefont {Volotka}}, \ and\ \bibinfo {author} {\bibfnamefont {G.}~\bibnamefont {Werth}},\ }\href {\doibase 10.1038/ncomms10246} {\bibfield  {journal} {\bibinfo  {journal} {Nature Communications}\ }\textbf {\bibinfo {volume} {7}},\ \bibinfo {pages} {10246} (\bibinfo {year} {2016})}\BibitemShut {NoStop}%
\bibitem [{\citenamefont {Arapoglou}\ \emph {et~al.}(2019)\citenamefont {Arapoglou}, \citenamefont {Egl}, \citenamefont {H{\"o}cker}, \citenamefont {Sailer}, \citenamefont {Tu}, \citenamefont {Weigel}, \citenamefont {Wolf}, \citenamefont {Cakir}, \citenamefont {Yerokhin}, \citenamefont {Oreshkina}, \citenamefont {Agababaev}, \citenamefont {Volotka}, \citenamefont {Zinenko}, \citenamefont {Glazov}, \citenamefont {Harman}, \citenamefont {Keitel}, \citenamefont {Sturm},\ and\ \citenamefont {Blaum}}]{Arapoglou:2019:253001}%
  \BibitemOpen
  \bibfield  {author} {\bibinfo {author} {\bibfnamefont {I.}~\bibnamefont {Arapoglou}}, \bibinfo {author} {\bibfnamefont {A.}~\bibnamefont {Egl}}, \bibinfo {author} {\bibfnamefont {M.}~\bibnamefont {H{\"o}cker}}, \bibinfo {author} {\bibfnamefont {T.}~\bibnamefont {Sailer}}, \bibinfo {author} {\bibfnamefont {B.}~\bibnamefont {Tu}}, \bibinfo {author} {\bibfnamefont {A.}~\bibnamefont {Weigel}}, \bibinfo {author} {\bibfnamefont {R.}~\bibnamefont {Wolf}}, \bibinfo {author} {\bibfnamefont {H.}~\bibnamefont {Cakir}}, \bibinfo {author} {\bibfnamefont {V.~A.}\ \bibnamefont {Yerokhin}}, \bibinfo {author} {\bibfnamefont {N.~S.}\ \bibnamefont {Oreshkina}}, \bibinfo {author} {\bibfnamefont {V.~A.}\ \bibnamefont {Agababaev}}, \bibinfo {author} {\bibfnamefont {A.~V.}\ \bibnamefont {Volotka}}, \bibinfo {author} {\bibfnamefont {D.~V.}\ \bibnamefont {Zinenko}}, \bibinfo {author} {\bibfnamefont {D.~A.}\ \bibnamefont {Glazov}}, \bibinfo {author} {\bibfnamefont {Z.}~\bibnamefont {Harman}}, \bibinfo {author} {\bibfnamefont
  {C.~H.}\ \bibnamefont {Keitel}}, \bibinfo {author} {\bibfnamefont {S.}~\bibnamefont {Sturm}}, \ and\ \bibinfo {author} {\bibfnamefont {K.}~\bibnamefont {Blaum}},\ }\href {\doibase 10.1103/PhysRevLett.122.253001} {\bibfield  {journal} {\bibinfo  {journal} {Physical Review Letters}\ }\textbf {\bibinfo {volume} {122}},\ \bibinfo {pages} {253001} (\bibinfo {year} {2019})}\BibitemShut {NoStop}%
\bibitem [{\citenamefont {Soria~Orts}\ \emph {et~al.}(2007)\citenamefont {Soria~Orts}, \citenamefont {{Crespo L{\'o}pez-Urrutia}}, \citenamefont {Bruhns}, \citenamefont {Gonz{\'a}lez~Mart{\'i}nez}, \citenamefont {Harman}, \citenamefont {Jentschura}, \citenamefont {Keitel}, \citenamefont {Lapierre}, \citenamefont {Tawara}, \citenamefont {Tupitsyn}, \citenamefont {Ullrich},\ and\ \citenamefont {Volotka}}]{SoriaOrts:2007:052501}%
  \BibitemOpen
  \bibfield  {author} {\bibinfo {author} {\bibfnamefont {R.}~\bibnamefont {Soria~Orts}}, \bibinfo {author} {\bibfnamefont {J.~R.}\ \bibnamefont {{Crespo L{\'o}pez-Urrutia}}}, \bibinfo {author} {\bibfnamefont {H.}~\bibnamefont {Bruhns}}, \bibinfo {author} {\bibfnamefont {A.~J.}\ \bibnamefont {Gonz{\'a}lez~Mart{\'i}nez}}, \bibinfo {author} {\bibfnamefont {Z.}~\bibnamefont {Harman}}, \bibinfo {author} {\bibfnamefont {U.~D.}\ \bibnamefont {Jentschura}}, \bibinfo {author} {\bibfnamefont {C.~H.}\ \bibnamefont {Keitel}}, \bibinfo {author} {\bibfnamefont {A.}~\bibnamefont {Lapierre}}, \bibinfo {author} {\bibfnamefont {H.}~\bibnamefont {Tawara}}, \bibinfo {author} {\bibfnamefont {I.~I.}\ \bibnamefont {Tupitsyn}}, \bibinfo {author} {\bibfnamefont {J.}~\bibnamefont {Ullrich}}, \ and\ \bibinfo {author} {\bibfnamefont {A.~V.}\ \bibnamefont {Volotka}},\ }\href {\doibase 10.1103/PhysRevA.76.052501} {\bibfield  {journal} {\bibinfo  {journal} {Physical Review A}\ }\textbf {\bibinfo {volume} {76}},\ \bibinfo {pages} {052501}
  (\bibinfo {year} {2007})}\BibitemShut {NoStop}%
\bibitem [{\citenamefont {Egl}\ \emph {et~al.}(2019)\citenamefont {Egl}, \citenamefont {Arapoglou}, \citenamefont {H{\"o}cker}, \citenamefont {K{\"o}nig}, \citenamefont {Ratajczyk}, \citenamefont {Sailer}, \citenamefont {Tu}, \citenamefont {Weigel}, \citenamefont {Blaum}, \citenamefont {N{\"o}rtersh{\"a}user},\ and\ \citenamefont {Sturm}}]{Egl:2019:123001}%
  \BibitemOpen
  \bibfield  {author} {\bibinfo {author} {\bibfnamefont {A.}~\bibnamefont {Egl}}, \bibinfo {author} {\bibfnamefont {I.}~\bibnamefont {Arapoglou}}, \bibinfo {author} {\bibfnamefont {M.}~\bibnamefont {H{\"o}cker}}, \bibinfo {author} {\bibfnamefont {K.}~\bibnamefont {K{\"o}nig}}, \bibinfo {author} {\bibfnamefont {T.}~\bibnamefont {Ratajczyk}}, \bibinfo {author} {\bibfnamefont {T.}~\bibnamefont {Sailer}}, \bibinfo {author} {\bibfnamefont {B.}~\bibnamefont {Tu}}, \bibinfo {author} {\bibfnamefont {A.}~\bibnamefont {Weigel}}, \bibinfo {author} {\bibfnamefont {K.}~\bibnamefont {Blaum}}, \bibinfo {author} {\bibfnamefont {W.}~\bibnamefont {N{\"o}rtersh{\"a}user}}, \ and\ \bibinfo {author} {\bibfnamefont {S.}~\bibnamefont {Sturm}},\ }\href {\doibase 10.1103/PhysRevLett.123.123001} {\bibfield  {journal} {\bibinfo  {journal} {Physical Review Letters}\ }\textbf {\bibinfo {volume} {123}},\ \bibinfo {pages} {123001} (\bibinfo {year} {2019})}\BibitemShut {NoStop}%
\bibitem [{\citenamefont {Micke}\ \emph {et~al.}(2020)\citenamefont {Micke}, \citenamefont {Leopold}, \citenamefont {King}, \citenamefont {Benkler}, \citenamefont {Spie{\ss}}, \citenamefont {Schm{\"o}ger}, \citenamefont {Schwarz}, \citenamefont {{Crespo L{\'o}pez-Urrutia}},\ and\ \citenamefont {Schmidt}}]{Micke:2020:60}%
  \BibitemOpen
  \bibfield  {author} {\bibinfo {author} {\bibfnamefont {P.}~\bibnamefont {Micke}}, \bibinfo {author} {\bibfnamefont {T.}~\bibnamefont {Leopold}}, \bibinfo {author} {\bibfnamefont {S.~A.}\ \bibnamefont {King}}, \bibinfo {author} {\bibfnamefont {E.}~\bibnamefont {Benkler}}, \bibinfo {author} {\bibfnamefont {L.~J.}\ \bibnamefont {Spie{\ss}}}, \bibinfo {author} {\bibfnamefont {L.}~\bibnamefont {Schm{\"o}ger}}, \bibinfo {author} {\bibfnamefont {M.}~\bibnamefont {Schwarz}}, \bibinfo {author} {\bibfnamefont {J.~R.}\ \bibnamefont {{Crespo L{\'o}pez-Urrutia}}}, \ and\ \bibinfo {author} {\bibfnamefont {P.~O.}\ \bibnamefont {Schmidt}},\ }\href {\doibase 10.1038/s41586-020-1959-8} {\bibfield  {journal} {\bibinfo  {journal} {Nature}\ }\textbf {\bibinfo {volume} {578}},\ \bibinfo {pages} {60} (\bibinfo {year} {2020})}\BibitemShut {NoStop}%
\bibitem [{\citenamefont {Spie{\ss}}\ \emph {et~al.}(2025)\citenamefont {Spie{\ss}}, \citenamefont {Chen}, \citenamefont {Wilzewski}, \citenamefont {Wehrheim}, \citenamefont {Gilles}, \citenamefont {Surzhykov}, \citenamefont {Benkler}, \citenamefont {Filzinger}, \citenamefont {Steinel}, \citenamefont {Huntemann}, \citenamefont {Cheung}, \citenamefont {Porsev}, \citenamefont {Bondarev}, \citenamefont {Safronova}, \citenamefont {{L{\'o}pez-Urrutia}},\ and\ \citenamefont {Schmidt}}]{spiess:25}%
  \BibitemOpen
  \bibfield  {author} {\bibinfo {author} {\bibfnamefont {L.~J.}\ \bibnamefont {Spie{\ss}}}, \bibinfo {author} {\bibfnamefont {S.}~\bibnamefont {Chen}}, \bibinfo {author} {\bibfnamefont {A.}~\bibnamefont {Wilzewski}}, \bibinfo {author} {\bibfnamefont {M.}~\bibnamefont {Wehrheim}}, \bibinfo {author} {\bibfnamefont {J.}~\bibnamefont {Gilles}}, \bibinfo {author} {\bibfnamefont {A.}~\bibnamefont {Surzhykov}}, \bibinfo {author} {\bibfnamefont {E.}~\bibnamefont {Benkler}}, \bibinfo {author} {\bibfnamefont {M.}~\bibnamefont {Filzinger}}, \bibinfo {author} {\bibfnamefont {M.}~\bibnamefont {Steinel}}, \bibinfo {author} {\bibfnamefont {N.}~\bibnamefont {Huntemann}}, \bibinfo {author} {\bibfnamefont {C.}~\bibnamefont {Cheung}}, \bibinfo {author} {\bibfnamefont {S.~G.}\ \bibnamefont {Porsev}}, \bibinfo {author} {\bibfnamefont {A.~I.}\ \bibnamefont {Bondarev}}, \bibinfo {author} {\bibfnamefont {M.~S.}\ \bibnamefont {Safronova}}, \bibinfo {author} {\bibfnamefont {J.~R.~C.}\ \bibnamefont {{L{\'o}pez-Urrutia}}}, \ and\
  \bibinfo {author} {\bibfnamefont {P.~O.}\ \bibnamefont {Schmidt}},\ }\href@noop {} {\bibfield  {journal} {\bibinfo  {journal} {{arXiv.2502.18926 [physics.atom-ph]}}\ } (\bibinfo {year} {2025})}\BibitemShut {NoStop}%
\bibitem [{\citenamefont {Guan}\ and\ \citenamefont {Wang}(1998)}]{Guan:1998:120}%
  \BibitemOpen
  \bibfield  {author} {\bibinfo {author} {\bibfnamefont {X.~X.}\ \bibnamefont {Guan}}\ and\ \bibinfo {author} {\bibfnamefont {Z.~W.}\ \bibnamefont {Wang}},\ }\href {\doibase 10.1016/S0375-9601(98)00332-6} {\bibfield  {journal} {\bibinfo  {journal} {Physics Letters A}\ }\textbf {\bibinfo {volume} {244}},\ \bibinfo {pages} {120} (\bibinfo {year} {1998})}\BibitemShut {NoStop}%
\bibitem [{\citenamefont {Yan}(2001)}]{Yan:2001:5683}%
  \BibitemOpen
  \bibfield  {author} {\bibinfo {author} {\bibfnamefont {Z.-C.}\ \bibnamefont {Yan}},\ }\href {\doibase 10.1103/PhysRevLett.86.5683} {\bibfield  {journal} {\bibinfo  {journal} {Physical Review Letters}\ }\textbf {\bibinfo {volume} {86}},\ \bibinfo {pages} {5683} (\bibinfo {year} {2001})}\BibitemShut {NoStop}%
\bibitem [{\citenamefont {Yan}(2002{\natexlab{a}})}]{Yan:2002:1885}%
  \BibitemOpen
  \bibfield  {author} {\bibinfo {author} {\bibfnamefont {Z.-C.}\ \bibnamefont {Yan}},\ }\href {\doibase 10.1088/0953-4075/35/8/307} {\bibfield  {journal} {\bibinfo  {journal} {Journal of Physics B: Atomic, Molecular and Optical Physics}\ }\textbf {\bibinfo {volume} {35}},\ \bibinfo {pages} {1885} (\bibinfo {year} {2002}{\natexlab{a}})}\BibitemShut {NoStop}%
\bibitem [{\citenamefont {Yan}(2002{\natexlab{b}})}]{Yan:2002:022502}%
  \BibitemOpen
  \bibfield  {author} {\bibinfo {author} {\bibfnamefont {Z.-C.}\ \bibnamefont {Yan}},\ }\href {\doibase 10.1103/PhysRevA.66.022502} {\bibfield  {journal} {\bibinfo  {journal} {Physical Review A}\ }\textbf {\bibinfo {volume} {66}},\ \bibinfo {pages} {022502} (\bibinfo {year} {2002}{\natexlab{b}})}\BibitemShut {NoStop}%
\bibitem [{\citenamefont {Glazov}\ \emph {et~al.}(2013)\citenamefont {Glazov}, \citenamefont {Volotka}, \citenamefont {Schepetnov}, \citenamefont {Sokolov}, \citenamefont {Shabaev}, \citenamefont {Tupitsyn},\ and\ \citenamefont {Plunien}}]{Glazov:2013:014014}%
  \BibitemOpen
  \bibfield  {author} {\bibinfo {author} {\bibfnamefont {D.~A.}\ \bibnamefont {Glazov}}, \bibinfo {author} {\bibfnamefont {A.~V.}\ \bibnamefont {Volotka}}, \bibinfo {author} {\bibfnamefont {A.~A.}\ \bibnamefont {Schepetnov}}, \bibinfo {author} {\bibfnamefont {M.~M.}\ \bibnamefont {Sokolov}}, \bibinfo {author} {\bibfnamefont {V.~M.}\ \bibnamefont {Shabaev}}, \bibinfo {author} {\bibfnamefont {I.~I.}\ \bibnamefont {Tupitsyn}}, \ and\ \bibinfo {author} {\bibfnamefont {G.}~\bibnamefont {Plunien}},\ }\href {\doibase 10.1088/0031-8949/2013/T156/014014} {\bibfield  {journal} {\bibinfo  {journal} {Physica Scripta}\ }\textbf {\bibinfo {volume} {T156}},\ \bibinfo {pages} {014014} (\bibinfo {year} {2013})}\BibitemShut {NoStop}%
\bibitem [{\citenamefont {Maison}\ \emph {et~al.}(2019)\citenamefont {Maison}, \citenamefont {Skripnikov},\ and\ \citenamefont {Glazov}}]{Maison:2019:042506}%
  \BibitemOpen
  \bibfield  {author} {\bibinfo {author} {\bibfnamefont {D.~E.}\ \bibnamefont {Maison}}, \bibinfo {author} {\bibfnamefont {L.~V.}\ \bibnamefont {Skripnikov}}, \ and\ \bibinfo {author} {\bibfnamefont {D.~A.}\ \bibnamefont {Glazov}},\ }\href {\doibase 10.1103/PhysRevA.99.042506} {\bibfield  {journal} {\bibinfo  {journal} {Physical Review A}\ }\textbf {\bibinfo {volume} {99}},\ \bibinfo {pages} {042506} (\bibinfo {year} {2019})}\BibitemShut {NoStop}%
\bibitem [{\citenamefont {Agababaev}\ \emph {et~al.}(2020)\citenamefont {Agababaev}, \citenamefont {Glazov}, \citenamefont {Volotka}, \citenamefont {Zinenko}, \citenamefont {Shabaev},\ and\ \citenamefont {Plunien}}]{Agababaev:2020:143}%
  \BibitemOpen
  \bibfield  {author} {\bibinfo {author} {\bibfnamefont {V.~A.}\ \bibnamefont {Agababaev}}, \bibinfo {author} {\bibfnamefont {D.~A.}\ \bibnamefont {Glazov}}, \bibinfo {author} {\bibfnamefont {A.~V.}\ \bibnamefont {Volotka}}, \bibinfo {author} {\bibfnamefont {D.~V.}\ \bibnamefont {Zinenko}}, \bibinfo {author} {\bibfnamefont {V.~M.}\ \bibnamefont {Shabaev}}, \ and\ \bibinfo {author} {\bibfnamefont {G.}~\bibnamefont {Plunien}},\ }\href {\doibase 10.1002/xrs.3074} {\bibfield  {journal} {\bibinfo  {journal} {X-Ray Spectrometry}\ }\textbf {\bibinfo {volume} {49}},\ \bibinfo {pages} {143} (\bibinfo {year} {2020})}\BibitemShut {NoStop}%
\bibitem [{\citenamefont {Morgner}\ \emph {et~al.}(2025)\citenamefont {Morgner}, \citenamefont {Tu}, \citenamefont {Moretti}, \citenamefont {K{\"o}nig}, \citenamefont {Hei{\ss}e}, \citenamefont {Sailer}, \citenamefont {Yerokhin}, \citenamefont {Sikora}, \citenamefont {Oreshkina}, \citenamefont {Harman}, \citenamefont {Keitel}, \citenamefont {Sturm},\ and\ \citenamefont {Blaum}}]{Morgner:2025:123201}%
  \BibitemOpen
  \bibfield  {author} {\bibinfo {author} {\bibfnamefont {J.}~\bibnamefont {Morgner}}, \bibinfo {author} {\bibfnamefont {B.}~\bibnamefont {Tu}}, \bibinfo {author} {\bibfnamefont {M.}~\bibnamefont {Moretti}}, \bibinfo {author} {\bibfnamefont {C.~M.}\ \bibnamefont {K{\"o}nig}}, \bibinfo {author} {\bibfnamefont {F.}~\bibnamefont {Hei{\ss}e}}, \bibinfo {author} {\bibfnamefont {T.}~\bibnamefont {Sailer}}, \bibinfo {author} {\bibfnamefont {V.~A.}\ \bibnamefont {Yerokhin}}, \bibinfo {author} {\bibfnamefont {B.}~\bibnamefont {Sikora}}, \bibinfo {author} {\bibfnamefont {N.~S.}\ \bibnamefont {Oreshkina}}, \bibinfo {author} {\bibfnamefont {Z.}~\bibnamefont {Harman}}, \bibinfo {author} {\bibfnamefont {C.~H.}\ \bibnamefont {Keitel}}, \bibinfo {author} {\bibfnamefont {S.}~\bibnamefont {Sturm}}, \ and\ \bibinfo {author} {\bibfnamefont {K.}~\bibnamefont {Blaum}},\ }\href {\doibase 10.1103/PhysRevLett.134.123201} {\bibfield  {journal} {\bibinfo  {journal} {Physical Review Letters}\ }\textbf {\bibinfo {volume} {134}},\ \bibinfo
  {pages} {123201} (\bibinfo {year} {2025})}\BibitemShut {NoStop}%
\bibitem [{\citenamefont {Verdebout}\ \emph {et~al.}(2014)\citenamefont {Verdebout}, \citenamefont {Naz{\'e}}, \citenamefont {J{\"o}nsson}, \citenamefont {Rynkun}, \citenamefont {Godefroid},\ and\ \citenamefont {Gaigalas}}]{Verdebout:2014:1111}%
  \BibitemOpen
  \bibfield  {author} {\bibinfo {author} {\bibfnamefont {S.}~\bibnamefont {Verdebout}}, \bibinfo {author} {\bibfnamefont {C.}~\bibnamefont {Naz{\'e}}}, \bibinfo {author} {\bibfnamefont {P.}~\bibnamefont {J{\"o}nsson}}, \bibinfo {author} {\bibfnamefont {P.}~\bibnamefont {Rynkun}}, \bibinfo {author} {\bibfnamefont {M.}~\bibnamefont {Godefroid}}, \ and\ \bibinfo {author} {\bibfnamefont {G.}~\bibnamefont {Gaigalas}},\ }\href {\doibase 10.1016/j.adt.2014.05.001} {\bibfield  {journal} {\bibinfo  {journal} {Atomic Data and Nuclear Data Tables}\ }\textbf {\bibinfo {volume} {100}},\ \bibinfo {pages} {1111} (\bibinfo {year} {2014})}\BibitemShut {NoStop}%
\bibitem [{\citenamefont {Marques}\ \emph {et~al.}(2016)\citenamefont {Marques}, \citenamefont {Indelicato}, \citenamefont {Parente}, \citenamefont {Sampaio},\ and\ \citenamefont {Santos}}]{Marques:2016:042504}%
  \BibitemOpen
  \bibfield  {author} {\bibinfo {author} {\bibfnamefont {J.~P.}\ \bibnamefont {Marques}}, \bibinfo {author} {\bibfnamefont {P.}~\bibnamefont {Indelicato}}, \bibinfo {author} {\bibfnamefont {F.}~\bibnamefont {Parente}}, \bibinfo {author} {\bibfnamefont {J.~M.}\ \bibnamefont {Sampaio}}, \ and\ \bibinfo {author} {\bibfnamefont {J.~P.}\ \bibnamefont {Santos}},\ }\href {\doibase 10.1103/PhysRevA.94.042504} {\bibfield  {journal} {\bibinfo  {journal} {Physical Review A}\ }\textbf {\bibinfo {volume} {94}},\ \bibinfo {pages} {042504} (\bibinfo {year} {2016})}\BibitemShut {NoStop}%
\bibitem [{\citenamefont {Sapirstein}\ and\ \citenamefont {Cheng}(2002)}]{Sapirstein:2002:042501}%
  \BibitemOpen
  \bibfield  {author} {\bibinfo {author} {\bibfnamefont {J.}~\bibnamefont {Sapirstein}}\ and\ \bibinfo {author} {\bibfnamefont {K.~T.}\ \bibnamefont {Cheng}},\ }\href {\doibase 10.1103/PhysRevA.66.042501} {\bibfield  {journal} {\bibinfo  {journal} {Physical Review A}\ }\textbf {\bibinfo {volume} {66}},\ \bibinfo {pages} {042501} (\bibinfo {year} {2002})}\BibitemShut {NoStop}%
\bibitem [{\citenamefont {{von Lindenfels}}\ \emph {et~al.}(2013)\citenamefont {{von Lindenfels}}, \citenamefont {Wiesel}, \citenamefont {Glazov}, \citenamefont {Volotka}, \citenamefont {Sokolov}, \citenamefont {Shabaev}, \citenamefont {Plunien}, \citenamefont {Quint}, \citenamefont {Birkl}, \citenamefont {Martin},\ and\ \citenamefont {Vogel}}]{vonLindenfels:2013:023412}%
  \BibitemOpen
  \bibfield  {author} {\bibinfo {author} {\bibfnamefont {D.}~\bibnamefont {{von Lindenfels}}}, \bibinfo {author} {\bibfnamefont {M.}~\bibnamefont {Wiesel}}, \bibinfo {author} {\bibfnamefont {D.~A.}\ \bibnamefont {Glazov}}, \bibinfo {author} {\bibfnamefont {A.~V.}\ \bibnamefont {Volotka}}, \bibinfo {author} {\bibfnamefont {M.~M.}\ \bibnamefont {Sokolov}}, \bibinfo {author} {\bibfnamefont {V.~M.}\ \bibnamefont {Shabaev}}, \bibinfo {author} {\bibfnamefont {G.}~\bibnamefont {Plunien}}, \bibinfo {author} {\bibfnamefont {W.}~\bibnamefont {Quint}}, \bibinfo {author} {\bibfnamefont {G.}~\bibnamefont {Birkl}}, \bibinfo {author} {\bibfnamefont {A.}~\bibnamefont {Martin}}, \ and\ \bibinfo {author} {\bibfnamefont {M.}~\bibnamefont {Vogel}},\ }\href {\doibase 10.1103/PhysRevA.87.023412} {\bibfield  {journal} {\bibinfo  {journal} {Physical Review A}\ }\textbf {\bibinfo {volume} {87}},\ \bibinfo {pages} {023412} (\bibinfo {year} {2013})}\BibitemShut {NoStop}%
\bibitem [{\citenamefont {Agababaev}\ \emph {et~al.}(2017)\citenamefont {Agababaev}, \citenamefont {Volchkova}, \citenamefont {Varentsova}, \citenamefont {Glazov}, \citenamefont {Volotka}, \citenamefont {Shabaev},\ and\ \citenamefont {Plunien}}]{Agababaev:17:NIMB}%
  \BibitemOpen
  \bibfield  {author} {\bibinfo {author} {\bibfnamefont {V.~A.}\ \bibnamefont {Agababaev}}, \bibinfo {author} {\bibfnamefont {A.~M.}\ \bibnamefont {Volchkova}}, \bibinfo {author} {\bibfnamefont {A.~S.}\ \bibnamefont {Varentsova}}, \bibinfo {author} {\bibfnamefont {D.~A.}\ \bibnamefont {Glazov}}, \bibinfo {author} {\bibfnamefont {A.~V.}\ \bibnamefont {Volotka}}, \bibinfo {author} {\bibfnamefont {V.~M.}\ \bibnamefont {Shabaev}}, \ and\ \bibinfo {author} {\bibfnamefont {G.}~\bibnamefont {Plunien}},\ }\href@noop {} {\bibfield  {journal} {\bibinfo  {journal} {Nucl. Intsr. Met. B}\ }\textbf {\bibinfo {volume} {408}},\ \bibinfo {pages} {70} (\bibinfo {year} {2017})}\BibitemShut {NoStop}%
\bibitem [{\citenamefont {Varentsova}\ \emph {et~al.}(2017)\citenamefont {Varentsova}, \citenamefont {Agababaev}, \citenamefont {Volchkova}, \citenamefont {Glazov}, \citenamefont {Volotka}, \citenamefont {Shabaev},\ and\ \citenamefont {Plunien}}]{Varentsova:2017:80}%
  \BibitemOpen
  \bibfield  {author} {\bibinfo {author} {\bibfnamefont {A.~S.}\ \bibnamefont {Varentsova}}, \bibinfo {author} {\bibfnamefont {V.~A.}\ \bibnamefont {Agababaev}}, \bibinfo {author} {\bibfnamefont {A.~M.}\ \bibnamefont {Volchkova}}, \bibinfo {author} {\bibfnamefont {D.~A.}\ \bibnamefont {Glazov}}, \bibinfo {author} {\bibfnamefont {A.~V.}\ \bibnamefont {Volotka}}, \bibinfo {author} {\bibfnamefont {V.~M.}\ \bibnamefont {Shabaev}}, \ and\ \bibinfo {author} {\bibfnamefont {G.}~\bibnamefont {Plunien}},\ }\href {\doibase 10.1016/j.nimb.2017.05.040} {\bibfield  {journal} {\bibinfo  {journal} {Nucl. Instrum. Methods Phys. Res. Sect. B}\ }\bibinfo {series} {Proceedings of the 18th {{International Conference}} on the {{Physics}} of {{Highly Charged Ions}} ({{HCI-2016}}), {{Kielce}}, {{Poland}}, 11-16 {{September}} 2016},\ \textbf {\bibinfo {volume} {408}},\ \bibinfo {pages} {80} (\bibinfo {year} {2017})}\BibitemShut {NoStop}%
\bibitem [{\citenamefont {Varentsova}\ \emph {et~al.}(2018)\citenamefont {Varentsova}, \citenamefont {Agababaev}, \citenamefont {Glazov}, \citenamefont {Volchkova}, \citenamefont {Volotka}, \citenamefont {Shabaev},\ and\ \citenamefont {Plunien}}]{Varentsova:2018:043402}%
  \BibitemOpen
  \bibfield  {author} {\bibinfo {author} {\bibfnamefont {A.~S.}\ \bibnamefont {Varentsova}}, \bibinfo {author} {\bibfnamefont {V.~A.}\ \bibnamefont {Agababaev}}, \bibinfo {author} {\bibfnamefont {D.~A.}\ \bibnamefont {Glazov}}, \bibinfo {author} {\bibfnamefont {A.~M.}\ \bibnamefont {Volchkova}}, \bibinfo {author} {\bibfnamefont {A.~V.}\ \bibnamefont {Volotka}}, \bibinfo {author} {\bibfnamefont {V.~M.}\ \bibnamefont {Shabaev}}, \ and\ \bibinfo {author} {\bibfnamefont {G.}~\bibnamefont {Plunien}},\ }\href {\doibase 10.1103/PhysRevA.97.043402} {\bibfield  {journal} {\bibinfo  {journal} {Phys. Rev. A}\ }\textbf {\bibinfo {volume} {97}},\ \bibinfo {pages} {043402} (\bibinfo {year} {2018})}\BibitemShut {NoStop}%
\bibitem [{\citenamefont {Agababaev}\ \emph {et~al.}(2025)\citenamefont {Agababaev}, \citenamefont {Prokhorchuk}, \citenamefont {Glazov}, \citenamefont {Malyshev}, \citenamefont {Shabaev},\ and\ \citenamefont {Volotka}}]{Agababaev:2025}%
  \BibitemOpen
  \bibfield  {author} {\bibinfo {author} {\bibfnamefont {V.~A.}\ \bibnamefont {Agababaev}}, \bibinfo {author} {\bibfnamefont {E.~A.}\ \bibnamefont {Prokhorchuk}}, \bibinfo {author} {\bibfnamefont {D.~A.}\ \bibnamefont {Glazov}}, \bibinfo {author} {\bibfnamefont {A.~V.}\ \bibnamefont {Malyshev}}, \bibinfo {author} {\bibfnamefont {V.~M.}\ \bibnamefont {Shabaev}}, \ and\ \bibinfo {author} {\bibfnamefont {A.~V.}\ \bibnamefont {Volotka}},\ }\href@noop {} {\bibfield  {journal} {\bibinfo  {journal} {arXiv:2025.xxxxx}\ } (\bibinfo {year} {2025})}\BibitemShut {NoStop}%
\bibitem [{\citenamefont {Breit}(1928)}]{Breit:1928:649}%
  \BibitemOpen
  \bibfield  {author} {\bibinfo {author} {\bibfnamefont {G.}~\bibnamefont {Breit}},\ }\href {\doibase 10.1038/122649a0} {\bibfield  {journal} {\bibinfo  {journal} {Nature}\ }\textbf {\bibinfo {volume} {122}},\ \bibinfo {pages} {649} (\bibinfo {year} {1928})}\BibitemShut {NoStop}%
\bibitem [{\citenamefont {Zinenko}\ \emph {et~al.}(2023)\citenamefont {Zinenko}, \citenamefont {Glazov}, \citenamefont {Kosheleva}, \citenamefont {Volotka},\ and\ \citenamefont {Fritzsche}}]{Zinenko:2023:032815}%
  \BibitemOpen
  \bibfield  {author} {\bibinfo {author} {\bibfnamefont {D.~V.}\ \bibnamefont {Zinenko}}, \bibinfo {author} {\bibfnamefont {D.~A.}\ \bibnamefont {Glazov}}, \bibinfo {author} {\bibfnamefont {V.~P.}\ \bibnamefont {Kosheleva}}, \bibinfo {author} {\bibfnamefont {A.~V.}\ \bibnamefont {Volotka}}, \ and\ \bibinfo {author} {\bibfnamefont {S.}~\bibnamefont {Fritzsche}},\ }\href {\doibase 10.1103/PhysRevA.107.032815} {\bibfield  {journal} {\bibinfo  {journal} {Physical Review A}\ }\textbf {\bibinfo {volume} {107}},\ \bibinfo {pages} {032815} (\bibinfo {year} {2023})}\BibitemShut {NoStop}%
\bibitem [{\citenamefont {Shabaev}(2002)}]{shabaev:2002:119}%
  \BibitemOpen
  \bibfield  {author} {\bibinfo {author} {\bibfnamefont {V.~M.}\ \bibnamefont {Shabaev}},\ }\href {\doibase 10.1016/S0370-1573(01)00024-2} {\bibfield  {journal} {\bibinfo  {journal} {Physics Reports}\ }\textbf {\bibinfo {volume} {356}},\ \bibinfo {pages} {119} (\bibinfo {year} {2002})}\BibitemShut {NoStop}%
\bibitem [{\citenamefont {Lindgren}\ \emph {et~al.}(2004)\citenamefont {Lindgren}, \citenamefont {Salomonson},\ and\ \citenamefont {{\AA}s{\'e}n}}]{lindgren:2004:161}%
  \BibitemOpen
  \bibfield  {author} {\bibinfo {author} {\bibfnamefont {I.}~\bibnamefont {Lindgren}}, \bibinfo {author} {\bibfnamefont {S.}~\bibnamefont {Salomonson}}, \ and\ \bibinfo {author} {\bibfnamefont {B.}~\bibnamefont {{\AA}s{\'e}n}},\ }\href {\doibase 10.1016/j.physrep.2003.09.004} {\bibfield  {journal} {\bibinfo  {journal} {Physics Reports}\ }\textbf {\bibinfo {volume} {389}},\ \bibinfo {pages} {161} (\bibinfo {year} {2004})}\BibitemShut {NoStop}%
\bibitem [{\citenamefont {Andreev}\ \emph {et~al.}(2008)\citenamefont {Andreev}, \citenamefont {Labzowsky}, \citenamefont {Plunien},\ and\ \citenamefont {Solovyev}}]{andreev:2008:135}%
  \BibitemOpen
  \bibfield  {author} {\bibinfo {author} {\bibfnamefont {O.~Y.}\ \bibnamefont {Andreev}}, \bibinfo {author} {\bibfnamefont {L.~N.}\ \bibnamefont {Labzowsky}}, \bibinfo {author} {\bibfnamefont {G.}~\bibnamefont {Plunien}}, \ and\ \bibinfo {author} {\bibfnamefont {D.~A.}\ \bibnamefont {Solovyev}},\ }\href {\doibase 10.1016/j.physrep.2007.10.003} {\bibfield  {journal} {\bibinfo  {journal} {Physics Reports}\ }\textbf {\bibinfo {volume} {455}},\ \bibinfo {pages} {135} (\bibinfo {year} {2008})}\BibitemShut {NoStop}%
\bibitem [{\citenamefont {Yerokhin}\ and\ \citenamefont {Jentschura}(2010)}]{Yerokhin:2010:012502}%
  \BibitemOpen
  \bibfield  {author} {\bibinfo {author} {\bibfnamefont {V.~A.}\ \bibnamefont {Yerokhin}}\ and\ \bibinfo {author} {\bibfnamefont {U.~D.}\ \bibnamefont {Jentschura}},\ }\href {\doibase 10.1103/PhysRevA.81.012502} {\bibfield  {journal} {\bibinfo  {journal} {Physical Review A}\ }\textbf {\bibinfo {volume} {81}},\ \bibinfo {pages} {012502} (\bibinfo {year} {2010})}\BibitemShut {NoStop}%
\bibitem [{\citenamefont {Glazov}\ \emph {et~al.}(2006)\citenamefont {Glazov}, \citenamefont {Volotka}, \citenamefont {Shabaev}, \citenamefont {Tupitsyn},\ and\ \citenamefont {Plunien}}]{Glazov:2006:330}%
  \BibitemOpen
  \bibfield  {author} {\bibinfo {author} {\bibfnamefont {D.~A.}\ \bibnamefont {Glazov}}, \bibinfo {author} {\bibfnamefont {A.~V.}\ \bibnamefont {Volotka}}, \bibinfo {author} {\bibfnamefont {V.~M.}\ \bibnamefont {Shabaev}}, \bibinfo {author} {\bibfnamefont {I.~I.}\ \bibnamefont {Tupitsyn}}, \ and\ \bibinfo {author} {\bibfnamefont {G.}~\bibnamefont {Plunien}},\ }\href {\doibase 10.1016/j.physleta.2006.04.056} {\bibfield  {journal} {\bibinfo  {journal} {Physics Letters A}\ }\textbf {\bibinfo {volume} {357}},\ \bibinfo {pages} {330} (\bibinfo {year} {2006})}\BibitemShut {NoStop}%
\bibitem [{\citenamefont {Volotka}\ \emph {et~al.}(2006)\citenamefont {Volotka}, \citenamefont {Glazov}, \citenamefont {Plunien}, \citenamefont {Shabaev},\ and\ \citenamefont {Tupitsyn}}]{Volotka:2006:293}%
  \BibitemOpen
  \bibfield  {author} {\bibinfo {author} {\bibfnamefont {A.~V.}\ \bibnamefont {Volotka}}, \bibinfo {author} {\bibfnamefont {D.~A.}\ \bibnamefont {Glazov}}, \bibinfo {author} {\bibfnamefont {G.}~\bibnamefont {Plunien}}, \bibinfo {author} {\bibfnamefont {V.~M.}\ \bibnamefont {Shabaev}}, \ and\ \bibinfo {author} {\bibfnamefont {I.~I.}\ \bibnamefont {Tupitsyn}},\ }\href {\doibase 10.1140/epjd/e2006-00048-8} {\bibfield  {journal} {\bibinfo  {journal} {The European Physical Journal D}\ }\textbf {\bibinfo {volume} {38}},\ \bibinfo {pages} {293} (\bibinfo {year} {2006})}\BibitemShut {NoStop}%
\bibitem [{\citenamefont {Shabaev}\ \emph {et~al.}(2004)\citenamefont {Shabaev}, \citenamefont {Tupitsyn}, \citenamefont {Yerokhin}, \citenamefont {Plunien},\ and\ \citenamefont {Soff}}]{Shabaev:2004:130405}%
  \BibitemOpen
  \bibfield  {author} {\bibinfo {author} {\bibfnamefont {V.~M.}\ \bibnamefont {Shabaev}}, \bibinfo {author} {\bibfnamefont {I.~I.}\ \bibnamefont {Tupitsyn}}, \bibinfo {author} {\bibfnamefont {V.~A.}\ \bibnamefont {Yerokhin}}, \bibinfo {author} {\bibfnamefont {G.}~\bibnamefont {Plunien}}, \ and\ \bibinfo {author} {\bibfnamefont {G.}~\bibnamefont {Soff}},\ }\href {\doibase 10.1103/PhysRevLett.93.130405} {\bibfield  {journal} {\bibinfo  {journal} {Physical Review Letters}\ }\textbf {\bibinfo {volume} {93}},\ \bibinfo {pages} {130405} (\bibinfo {year} {2004})}\BibitemShut {NoStop}%
\bibitem [{\citenamefont {Yerokhin}\ \emph {et~al.}(2004)\citenamefont {Yerokhin}, \citenamefont {Indelicato},\ and\ \citenamefont {Shabaev}}]{Yerokhin:2004:052503}%
  \BibitemOpen
  \bibfield  {author} {\bibinfo {author} {\bibfnamefont {V.~A.}\ \bibnamefont {Yerokhin}}, \bibinfo {author} {\bibfnamefont {P.}~\bibnamefont {Indelicato}}, \ and\ \bibinfo {author} {\bibfnamefont {V.~M.}\ \bibnamefont {Shabaev}},\ }\href {\doibase 10.1103/PhysRevA.69.052503} {\bibfield  {journal} {\bibinfo  {journal} {Physical Review A}\ }\textbf {\bibinfo {volume} {69}},\ \bibinfo {pages} {052503} (\bibinfo {year} {2004})}\BibitemShut {NoStop}%
\bibitem [{\citenamefont {Yerokhin}\ and\ \citenamefont {Shabaev}(1999)}]{Yerokhin:1999:800}%
  \BibitemOpen
  \bibfield  {author} {\bibinfo {author} {\bibfnamefont {V.~A.}\ \bibnamefont {Yerokhin}}\ and\ \bibinfo {author} {\bibfnamefont {V.~M.}\ \bibnamefont {Shabaev}},\ }\href {\doibase 10.1103/PhysRevA.60.800} {\bibfield  {journal} {\bibinfo  {journal} {Phys. Rev. A}\ }\textbf {\bibinfo {volume} {60}},\ \bibinfo {pages} {800} (\bibinfo {year} {1999})}\BibitemShut {NoStop}%
\bibitem [{\citenamefont {Lee}\ \emph {et~al.}(2007)\citenamefont {Lee}, \citenamefont {Milstein}, \citenamefont {Terekhov},\ and\ \citenamefont {Karshenboim}}]{Lee:07:CJP}%
  \BibitemOpen
  \bibfield  {author} {\bibinfo {author} {\bibfnamefont {R.~N.}\ \bibnamefont {Lee}}, \bibinfo {author} {\bibfnamefont {A.~I.}\ \bibnamefont {Milstein}}, \bibinfo {author} {\bibfnamefont {I.~S.}\ \bibnamefont {Terekhov}}, \ and\ \bibinfo {author} {\bibfnamefont {S.~G.}\ \bibnamefont {Karshenboim}},\ }\href {\doibase 10.1139/p07-024} {\bibfield  {journal} {\bibinfo  {journal} {Canadian Journal of Physics}\ }\textbf {\bibinfo {volume} {85}},\ \bibinfo {pages} {541} (\bibinfo {year} {2007})}\BibitemShut {NoStop}%
\bibitem [{\citenamefont {Glazov}\ \emph {et~al.}(2004)\citenamefont {Glazov}, \citenamefont {Shabaev}, \citenamefont {Tupitsyn}, \citenamefont {Volotka}, \citenamefont {Yerokhin}, \citenamefont {Plunien},\ and\ \citenamefont {Soff}}]{Glazov:2004:062104}%
  \BibitemOpen
  \bibfield  {author} {\bibinfo {author} {\bibfnamefont {D.~A.}\ \bibnamefont {Glazov}}, \bibinfo {author} {\bibfnamefont {V.~M.}\ \bibnamefont {Shabaev}}, \bibinfo {author} {\bibfnamefont {I.~I.}\ \bibnamefont {Tupitsyn}}, \bibinfo {author} {\bibfnamefont {A.~V.}\ \bibnamefont {Volotka}}, \bibinfo {author} {\bibfnamefont {V.~A.}\ \bibnamefont {Yerokhin}}, \bibinfo {author} {\bibfnamefont {G.}~\bibnamefont {Plunien}}, \ and\ \bibinfo {author} {\bibfnamefont {G.}~\bibnamefont {Soff}},\ }\href {\doibase 10.1103/PhysRevA.70.062104} {\bibfield  {journal} {\bibinfo  {journal} {Physical Review A}\ }\textbf {\bibinfo {volume} {70}},\ \bibinfo {pages} {062104} (\bibinfo {year} {2004})}\BibitemShut {NoStop}%
\bibitem [{\citenamefont {Agababaev}\ \emph {et~al.}(2018)\citenamefont {Agababaev}, \citenamefont {Glazov}, \citenamefont {Volotka}, \citenamefont {Zinenko}, \citenamefont {Shabaev},\ and\ \citenamefont {Plunien}}]{Agababaev:2018:012003}%
  \BibitemOpen
  \bibfield  {author} {\bibinfo {author} {\bibfnamefont {V.~A.}\ \bibnamefont {Agababaev}}, \bibinfo {author} {\bibfnamefont {D.~A.}\ \bibnamefont {Glazov}}, \bibinfo {author} {\bibfnamefont {A.~V.}\ \bibnamefont {Volotka}}, \bibinfo {author} {\bibfnamefont {D.~V.}\ \bibnamefont {Zinenko}}, \bibinfo {author} {\bibfnamefont {V.~M.}\ \bibnamefont {Shabaev}}, \ and\ \bibinfo {author} {\bibfnamefont {G.}~\bibnamefont {Plunien}},\ }\href {\doibase 10.1088/1742-6596/1138/1/012003} {\bibfield  {journal} {\bibinfo  {journal} {Journal of Physics: Conference Series}\ }\textbf {\bibinfo {volume} {1138}},\ \bibinfo {pages} {012003} (\bibinfo {year} {2018})}\BibitemShut {NoStop}%
\bibitem [{\citenamefont {Pachucki}\ \emph {et~al.}(2005)\citenamefont {Pachucki}, \citenamefont {Czarnecki}, \citenamefont {Jentschura},\ and\ \citenamefont {Yerokhin}}]{Pachucki:2005:022108}%
  \BibitemOpen
  \bibfield  {author} {\bibinfo {author} {\bibfnamefont {K.}~\bibnamefont {Pachucki}}, \bibinfo {author} {\bibfnamefont {A.}~\bibnamefont {Czarnecki}}, \bibinfo {author} {\bibfnamefont {U.~D.}\ \bibnamefont {Jentschura}}, \ and\ \bibinfo {author} {\bibfnamefont {V.~A.}\ \bibnamefont {Yerokhin}},\ }\href {\doibase 10.1103/PhysRevA.72.022108} {\bibfield  {journal} {\bibinfo  {journal} {Physical Review A}\ }\textbf {\bibinfo {volume} {72}},\ \bibinfo {pages} {022108} (\bibinfo {year} {2005})}\BibitemShut {NoStop}%
\bibitem [{\citenamefont {Czarnecki}\ \emph {et~al.}(2020)\citenamefont {Czarnecki}, \citenamefont {Piclum},\ and\ \citenamefont {Szafron}}]{Czarnecki:2020:050801}%
  \BibitemOpen
  \bibfield  {author} {\bibinfo {author} {\bibfnamefont {A.}~\bibnamefont {Czarnecki}}, \bibinfo {author} {\bibfnamefont {J.}~\bibnamefont {Piclum}}, \ and\ \bibinfo {author} {\bibfnamefont {R.}~\bibnamefont {Szafron}},\ }\href {\doibase 10.1103/PhysRevA.102.050801} {\bibfield  {journal} {\bibinfo  {journal} {Physical Review A}\ }\textbf {\bibinfo {volume} {102}},\ \bibinfo {pages} {050801} (\bibinfo {year} {2020})}\BibitemShut {NoStop}%
\bibitem [{\citenamefont {Jentschura}(2010)}]{Jentschura:2010:012512}%
  \BibitemOpen
  \bibfield  {author} {\bibinfo {author} {\bibfnamefont {U.~D.}\ \bibnamefont {Jentschura}},\ }\href {\doibase 10.1103/PhysRevA.81.012512} {\bibfield  {journal} {\bibinfo  {journal} {Physical Review A}\ }\textbf {\bibinfo {volume} {81}},\ \bibinfo {pages} {012512} (\bibinfo {year} {2010})}\BibitemShut {NoStop}%
\bibitem [{\citenamefont {Shabaev}(2001)}]{Shabaev:2001:052104}%
  \BibitemOpen
  \bibfield  {author} {\bibinfo {author} {\bibfnamefont {V.~M.}\ \bibnamefont {Shabaev}},\ }\href {\doibase 10.1103/PhysRevA.64.052104} {\bibfield  {journal} {\bibinfo  {journal} {Phys. Rev. A}\ }\textbf {\bibinfo {volume} {64}},\ \bibinfo {pages} {052104} (\bibinfo {year} {2001})}\BibitemShut {NoStop}%
\bibitem [{\citenamefont {Malyshev}\ \emph {et~al.}(2020)\citenamefont {Malyshev}, \citenamefont {Glazov},\ and\ \citenamefont {Shabaev}}]{Malyshev:2020:012513}%
  \BibitemOpen
  \bibfield  {author} {\bibinfo {author} {\bibfnamefont {A.~V.}\ \bibnamefont {Malyshev}}, \bibinfo {author} {\bibfnamefont {D.~A.}\ \bibnamefont {Glazov}}, \ and\ \bibinfo {author} {\bibfnamefont {V.~M.}\ \bibnamefont {Shabaev}},\ }\href {\doibase 10.1103/PhysRevA.101.012513} {\bibfield  {journal} {\bibinfo  {journal} {Physical Review A}\ }\textbf {\bibinfo {volume} {101}},\ \bibinfo {pages} {012513} (\bibinfo {year} {2020})}\BibitemShut {NoStop}%
\bibitem [{\citenamefont {Sapirstein}\ and\ \citenamefont {Johnson}(1996)}]{Sapirstein:1996:5213}%
  \BibitemOpen
  \bibfield  {author} {\bibinfo {author} {\bibfnamefont {J.}~\bibnamefont {Sapirstein}}\ and\ \bibinfo {author} {\bibfnamefont {W.~R.}\ \bibnamefont {Johnson}},\ }\href {\doibase 10.1088/0953-4075/29/22/005} {\bibfield  {journal} {\bibinfo  {journal} {Journal of Physics B: Atomic, Molecular and Optical Physics}\ }\textbf {\bibinfo {volume} {29}},\ \bibinfo {pages} {5213} (\bibinfo {year} {1996})}\BibitemShut {NoStop}%
\end{thebibliography}%
%
\end{document}